\documentclass[usenatbib]{mn2e}
\usepackage{times}
\usepackage{graphicx}
\usepackage[ansinew]{inputenc}
\usepackage[T1]{fontenc}
\usepackage{lmodern}
\usepackage{amssymb,amsmath}
\usepackage{rotating}
\usepackage{multirow}
\numberwithin{equation}{section}

\title[Calibration of RAVE distances]{Calibration of RAVE distances to a large sample of Hipparcos stars}
\author[Charles Francis]{Charles Francis$^{1}$\thanks{E-mail: C.E.H.Francis.75@cantab.net} \\
$^{1}$25 Elphinstone Rd., Hastings, TN34 2EG, UK.}

\begin{document}

\pagerange{\pageref{firstpage}--\pageref{lastpage}} \pubyear{2013}

\maketitle

\label{firstpage}

\begin{abstract}
A magnitude limited population of 18\,808 Hipparcos stars is used to calibrate distances for 52\,794 RAVE stars, including dwarfs, giants, and pre-main sequence stars. I give treatments for a number of types of bias affecting calculation, including bias from the non-linear relationship between the quantity of interest (e.g., distance or distance modulus) and the measured quantity (parallax or visual magnitude), the Lutz-Kelker bias, and bias due to variation in density of the stellar population. The use of a magnitude bound minimises the Malmquist and the Lutz-Kelker bias, and avoids a measurement bias resulting from the greater accuracy of Hipparcos parallaxes for brighter stars. The calibration is applicable to stars in 2MASS when there is some way to determine stellar class with reasonable confidence. For RAVE this is possible for hot dwarfs and using $\log g $. The accuracy of the calibration is tested against Hipparcos stars with better than 2\% parallax errors, and by comparison of the RAVE velocity distribution with that of Hipparcos, and is found to improve upon previous estimates of luminosity distance. An estimate of the LSR from RAVE data, $ (U_{0}, V_{0}, W_{0}) = (14.9 \pm 1.7, 15.3 \pm 0.4, 6.9 \pm 0.1) $ km\,s$^{-1}$, shows excellent agreement with the current best estimate from XHIP. The RAVE velocity distribution confirms the alignment of stellar motions with spiral structure.
\end{abstract}


\section{Introduction}\label{sec:1}
\subsection{The calibration of luminosity distance}\label{sec:1.1}
The determination of luminosity distance is vital to studies of stellar kinematics outside the immediate neighbourhood of the Sun, to the determination of cluster distances which are an important step on the cosmological distance scale, and to studies of Galactic structure. Luminosity distance has traditionally been determined by calibration to nearby cluster stars, typically the Hyades, since cluster distance can be determined with greater accuracy than the parallax distance to individual stars. Issues arise from a variety of biases affecting the calculation and because both age and composition affect position on a colour-magnitude diagram (CMD).

Bilir et al. (2008) have given a calibration of the main sequence based on Hipparcos dwarfs with better than 5\% parallax errors. The present work advances that of Bilir et al. (2008) by dividing the main sequence into colour bands so as to model it more accurately, by testing different combinations of filter bands, finding the most accurate distance estimates by using absolute $H$ magnitude, $ M_{H} $, with $ J - K $ colour, by using a larger, magnitude limited, Hipparcos population to minimise the Malmquist bias and Lutz-Kelker bias and to avoid to a measurement bias which arises because Hipparcos parallaxes are more accurate for brighter stars, by treating other biases arising from non-linear relationships between astronomical quantities, and by calibrating distances for a number of populations of stars in addition to those on the main sequence.

More recently, a number of studies (Breddels et al. 2010; Zwitter, Matijevi\v{c} \& Breddels 2010; Burnett et al. 2011) have calculated luminosity distances by methods dependent on stellar evolutionary theory. The accuracy of these methods depends on the quality of the match between theoretical models and physical evolution, on the accuracy of measurement of stellar parameters such as surface gravity, metallicity and effective temperature, and also, critically, on the shape and density of isochrones on a CMD, which determines the degree of uncertainty with which an isochrone may be assigned to any given star. The problems in identification of isochrones due to systematic errors in stellar parameters calculated from RAVE spectra (Burnett et al. 2011) will be avoided through the calibration to Hipparcos stars, since only 2MASS magnitudes and $\log g$ are used in the determination of stellar class. A systematic error in $\log g$ would affect the vertical scale of the lower diagram of figure 4, but would have little effect on class membership.

The computer modelling of stellar evolution based on known fundamental processes in nuclear physics has been one of the most important successes in theoretical science in the last fifty years. Nonetheless, we cannot expect perfect correspondence between computer models and physical reality. In practice, measurements of effective temperature, surface gravity and metallicity are subject to significant errors, even for high resolution measurements (e.g. Anderson \& Francis 2012, hereafter XHIP, and references therein). Errors in RAVE metallicities are about an order of magnitude greater (Zwitter, Siebert \& Munari 2008). Isochrone ages are similarly subject to considerable uncertainty. The Hyades cluster is thought to have an age of about 625 Myr (Perryman et al. 1998), but for 83 member stars with isochrone ages given by Holmberg, Nordstr\"om and Andersen (2007), half have ages over 2.6 Gyr, the maximum is 12 Gyr, the mean is 3.4 Gyr and the standard deviation is 2.9 Gyr. `Isochrone' takes its meaning from the Greek, \textit{iso}, the same, and \textit{chronos}, time. An isochrone on a CMD is a line of equal age. The determination of a stellar isochrone is thus synonymous with determination of age. The identification of isochrones is possible for stellar clusters and may complement other determinations of cluster distances, but generally for field stars age affects position on a CMD, so that uncertainties in age contribute directly to uncertainty in luminosity distance. It is therefore important to complement luminosity distances calculated from isochrones with those based on empirical study.

This paper describes the calibration of five magnitude limited populations of stars from the Hipparcos new reduction (van Leeuwen 2007). Comparison with Hipparcos stars with accurate parallaxes shows smaller errors than previous studies. The calibration is applied to the RAVE third data release (Siebert et al. 2011), and leads to a data base of distances for 39\,514 RAVE dwarfs and 13\,280 stars in other classes. 24\,861 dwarfs and 12\,712 stars in other classes have usable radial velocities and proper motions. The validity of this data base is shown by comparison of the velocity distributions for RAVE and Hipparcos. Although Hipparcos is an all sky survey, and RAVE concentrated on stars in the Southern hemisphere away from the Galactic plane, both contain a full range of thin and thick disc stars. The calibration will not be adversely affected by differences in composition because it was found that metallicity has little effect on distances calculated using the $ J - K $ colour index (section 4.1), and that random variations in metallicity are much greater than systematic changes, even for stars in the thick disc (section \ref{sec:5.3}). 

The kinematic data for RAVE are clearly less accurate than those available for Hipparcos, because distances are less accurate, and because smaller and proportionately less accurate proper motions arise from the greater distances of RAVE stars. Nonetheless the population is of sufficient size that it is possible to study general features and identify features corresponding to those found in the Hipparcos velocity distribution, and, because RAVE stars are more distant than Hipparcos stars, it is possible to observe changes in the velocity distribution with distance from the Sun. 

When compared to 618 Hipparcos dwarfs with parallax errors less than 2\%, the derived luminosity distances show a net error of 19.9\%. This may be taken as a reasonable error estimate, because in a population of 16\,820 single dwarfs the removal of 618 stars, even those with the most accurate measurement, has very little effect on the statistical properties. For 639 T-Tauri and subgiants (`TT\&S'), the net error is 13.6\%, for 768 protostars and red giant branch (`PS\&R') it is 11.6\%, and for 570 red clump stars `RC', it is 10.4\%. For RAVE stars with Galactic latitude $ |b| >9.7 $\textdegree, corrections to distances for reddening and for stellar density are typically less than 5\% and contribute very little to errors. Quoted errors in 2MASS magnitudes (Cutri et al. 2003; Skrutskie et al. 2006) generate distance errors of the order of one percent, and make a very small contribution to the net errors. There is no other reason to increase the percentage error estimates for RAVE stars, but uncertainties in identification will lead to substantial errors for small numbers of stars. These figures show a worthwhile improvement on the errors reported by Zwitter et al. (2010), $ \sim$21\% for both dwarfs and giants, and those calculated for dwarfs using the method of Bilir et al. (2008), $ 21.4 $\%.

The improvement is actually better than is first apparent; if one calibrates dwarfs to a distance limited sample (as did Bilir et al. 2008) errors are better than 17\%. However, this does not take account of the Malmquist bias. Zwitter et al. (2010) also did not use a magnitude limited calibration sample, but instead calibrated to a small population of Hipparcos stars, selected for calibration of radial velocities. This may not be ideal for calibration of distances and will create an unknown bias, similar to the Malmquist or Lutz-Kelker bias (section 2), depending on the selection. In practice, distances for stars which appear in both data bases are on average 28\% less than those of Zwitter et al. (2010) for 8\,748 dwarfs and 26\% less for 5\,355 members of other classes. The difference owes as much to errors in the identification of stellar class as to calibration. By contrast, distances for RAVE dwarfs calculated by the method of Bilir et al. (2008) are on average 15\% less than those found here, probably mainly because Bilir et al. (2008) did not consider the Malmquist bias. 

\subsection{Paper structure}\label{sec:1.2}
For an accurate calibration, it is necessary to consider different types of bias which affect parallax and luminosity distances, as discussed in section 2. Section 3 compares the CMD for Hipparcos with the plot of $\log g$ against $ J - K $ for RAVE stars. It is found possible to identify five populations, main sequence, red clump, T-Tauri and subgiants, protostars and red giant branch, and white dwarfs with sufficient numbers and with sufficiently certain absolute magnitudes for the calibration of luminosity distance. With the exception of white dwarfs (whose surface gravities are so high that determination of radial velocity is problematic), corresponding populations can be identified in RAVE by surface gravity and colour.

Section 4.1 describes the calibration of luminosity distance to parallax distances for a magnitude limited population of 16 820 single dwarfs (the `calibration sample') with parallax errors better than 20\% and colours in the range $ -0.1 < J - K \le 0.75 $, using 2MASS $J$, $H$, and $K$ magnitudes for stars with no photometric quality flags raised (i.e. the flag is `AAA'). Components of multiple stars were removed using the Catalog of Components of Double \& Multiple Stars (Dommanget \& Nys 2002) and The Washington Visual Double Star Catalog (`WDS', Mason et al. 2001-2010). The calibration is repeated for other classes in section 4.2. 

Section 5.1 describes the calculation of distances for 52 794 RAVE stars with Galactic latitude $ |b| > 9.7 $\textdegree, using corrections for reddening. This represents a substantial increase in the number of RAVE stars with publicly available estimates of luminosity distance (data to be submitted to CDS). Section 5.2 calculates Solar motion perpendicular to the Galactic plane. Section 5.3 sets a pragmatic boundary between the thin and thick disc, and shows from the metallicity gradient that there is no sharp distinction between them, but rather a continuous progression from stars with lower vertical motions and higher metallicity towards greater vertical motion and lower metallicity, in accordance with the finding of Bovy et al. (2012) from SEGUE data. 

Section 5.4 discusses the velocity distributions of the thin disc populations with different ages finding that, with the exception of the Pleiades stream which is greatly reduced or displaced, the major stellar streams seen in the Solar neighbourhood in the kinematics of Hipparcos stars continue, with slow variation, to greater distances. In contrast to the distributions of thin disc dwarfs and giants, which have a well at the position of circular motion, there is a large peak close to this position in the distribution of new stars. The velocity distribution for thick disc giants is seen is section 5.5, and has very little structure.

Section 5.6 makes comparison with the velocity distributions of Hipparcos stars and RAVE stars using distances from Padova isochrones calculated by Zwitter et al. (2010). The population of RAVE stars only contains a few hundred Hipparcos stars, and no more than a couple of hundred with good parallaxes, so these populations are essentially independent. The distance of RAVE dwarfs is greater than that of Hipparcos stars, but not so distant that a substantial change in the velocity distribution is expected. Similar features are seen, but a better match with the Hipparcos velocity distribution is obtained using luminosity distances calibrated empirically to Hipparcos, than theoretically from stellar models. This may be taken as evidence that the empirical calibration of distance is more accurate than theoretical determination from stellar models. 

Section 5.7 repeats the calculation of the local standard of rest originally described by Francis and Anderson (2009a), finding $ (U_{0}, V_{0}, W_{0}) = (14.9 \pm 1.7, 15.3 \pm 0.4, 6.9 \pm 0.1) $ km\,s$^{-1}$ in close agreement with the Hipparcos value, $ (U_{0}, V_{0}, W_{0}) = (14.1 \pm 1.1, 14.6 \pm 0.4, 6.9 \pm 0.1)$ km\,s$^{-1}$, found from an almost entirely distinct population of stars. Section 5.8 discusses Galactic spiral structure, as revealed in the RAVE velocity distribution, finding very similar results to those found in Hipparcos stars. Conclusions are summarised in section 6.

\section{Causes of bias}\label{sec:2}
\subsection{Bias in measurement of distance}\label{sec:2.1}
In physical measurement we do not determine the actual value of a quantity, but rather state the quantity together with an error. In its general form, due to Lindeberg (1922), the central limit theorem states that, under reasonable conditions, the sum of a large set of independent random variables, each having an arbitrary probability distribution, can be approximated by a normal distribution. In practice, we may reasonably assume that the distribution of data influenced by many small and unrelated random effects is approximately normal. As this broadly true for physical measurements, we may usually regard the measured value and stated error as the mean and standard deviation of a normal, or Gaussian, probability density function (pdf) for the true value of the measured quantity.

A statistic is biased if it is calculated in such a way that is systematically different from the population parameter of interest. Bias is the difference between an estimator's expectation and the true expectation of the parameter being estimated. Sometimes it is more appropriate to talk of the bias factor, the ratio between an estimator's expectation and the true expectation. For example, in the calculation of luminosity distances, bias in magnitudes corresponds to bias factor in distances.

We may assume that parallax measurements, taken, for example, by Hipparcos, are normally distributed with the stated mean and error, but the quantity of interest, astronomical distance, is a non-linear function, $ R = 1000 / \pi $, of measured parallax, $ \pi $. In this case the distance calculated from the mean parallax is not equal to the mean of the distance probability distribution. It follows that the standard formula, $ R = 1000 / \pi $, is a biased estimator of astronomical distance. This will be corrected in section 2.3.

For luminosity distances, a number of other causes of bias enter calculation. For statistical studies in stellar kinematics, if we are not to distort the calculated velocity distribution, it is important that we use an unbiased estimator of astronomical distance by analysing and correcting the various causes of bias. The correction does not give the actual value of an astronomical distance (which is never known), but should ensure that the calculated value is the mean of the probability distribution for the distance found from the measurements.

Papers on the calculation of stellar distance often display formulae relating the probability density functions of various quantities. For example, Burnett \& Binney (2010) give a Bayesian formula (their eq. (5)) which expresses the pdf of true stellar characteristics (including distance) given particular measurement values, in terms of the selection function of the population under study, the pdf, assumed normal, of a measurement result given true values, and the pdf of the actual values, taken as a prior. However, the pdf for stellar distance is not normal. The selection function for RAVE depends upon magnitude, but luminosity distances are calculated on a star by star basis, given 2MASS magnitudes measured with high precision so that the selection function has no practical bearing on the result. This paper takes a simpler approach by considering relationships between mean values instead of relationships between distributions. Whereas Burnett \& Binney (2010), following a Bayesian approach, produce posterior pdfs of what they believe they know about a star's distance given some data, here (irrespective of philosophical differences between Bayesians and frequentists) I seek to give the best estimate of the object's distance in the form of an estimate of the mean value, corrected for bias, together with error bounds.

\subsection{Types of bias}\label{sec:2.2}
I consider three categories in the causes of bias. First, in a calibration, there may be a selection bias such that the characteristics of the target population are different from those of the calibration sample. This will cause an error in calibration. The Malmquist bias (section 2.6) and the Lutz-Kelker bias (section 2.8) are caused by selection bias. In principle calibration bias can be removed by multiplying by an appropriate factor, but in practice it may not be straightforward to calculate the appropriate factor and it is better to choose the calibration sample to be as similar as possible to the target population so as to avoid selection bias. Calibration bias can also arise if the quantity of interest has itself been found using a biased estimator on the calibration sample. This can happen for both the reasons considered below. 

Second, suppose the quantity of interest, $y$, is a non-linear function of the measured quantity, $ x $, for which the pdf is $ f(x) $ and the expected value is $ \overline{x}=E(x) $. Then, using the Taylor expansion, the expected value of $y$ is
\begin{equation}\label{eq:2.2.1}
\begin{split}
\overline{y}&=\int\limits_{-x}^{x}y(x)f(x)dx\\
&=y(\overline{x}) + y''\frac{(\overline{x})}{2!}\sigma^2 + \sum\limits_{n=3}^{\infty}\frac{y^{(n)}(\overline{x})}{n!}\mu_n,
\end{split}
\end{equation}
where $ \mu_n $ is the $ n $th central moment of the probability distribution (appendix A). Clearly $ \tilde{y} = y(\overline{x})$ is a biased estimator for $ \overline{y} $. This form of bias will be seen in parallax bias (section \ref{sec:2.3}), modulus bias (section \ref{sec:2.4}) and magnitude bias (section \ref{sec:2.5}). 

Third, bias will arise if the true distribution $ \rho(y) $ is not uniform. If $y$ is a non-linear function of the measured quantity, $ x $, and if $ f(x) $ is normal (as we expect for a measured quantity), then the pdf $ g(y) $ for the quantity of interest given the measurement results is not normal. We assume that $ g(y) = N(y,\overline{y}, \sigma_y) +h(y) $, where $ h(y) $ is small and 
\begin{equation}\label{eq:2.2.2}
\int\limits_{-\infty}^{\infty}yh(y)dy =0.
\end{equation}
If, as is usually the case, the measurement is independent of the actual distribution, then the full pdf is the product of $ \rho(y) $ with $ g(y) $:
\begin{equation}\label{eq:2.2.3}
\rho(y)g(y)=\rho(y)(N(y,\overline{y},\sigma_y) + h(y))
\end{equation}
and the expectation of $y$ is
\begin{equation}\label{eq:2.2.4}
\begin{split}
E(y) &= \int\limits_{-\infty}^{\infty}y\rho(y)g(y)dy\\
&= \int\limits_{-\infty}^{\infty} y\rho(y)(N(y,\overline{y},\sigma_y) + h(y)) dy.
\end{split}
\end{equation}
If we can approximate $\rho$ with a linear function within $ 3\sigma_y $ of $ \overline{y} $,
\begin{equation}\label{eq:2.2.5}
\rho(y) \approx 1+ \rho'(y-\overline{y})
\end{equation}
where $\rho'$ is constant, then, using equation (\ref{eq:2.2.2}), we have 
\begin{equation}\label{eq:2.2.6}
E(y) \approx \overline{y} + \rho' \int\limits_{-\infty}^{\infty} y(y-\overline{y})(N(y,\overline{y},\sigma_y) + h(y)) dy.
\end{equation}
If $\rho$ is slowly varying ($\rho'$ is small) then the term containing $ h $ is a second order correction which can be ignored. Let $ z=(y-\overline{y})/\sigma_y $. Then
\begin{equation}\label{eq:2.2.8}
\begin{split}
E(y) &\approx \overline{y} +\sigma_y\rho' \int\limits_{-\infty}^{\infty} z (\overline{y}+ z\sigma_y)N(z,0,1)dz\\
&= \overline{y} +\sigma_y^2\rho' \int\limits_{-\infty}^{\infty} z^2 N(z,0,1) dz,
\end{split}
\end{equation}
since $ \overline{z} = 0 $. Then, using integration by parts
\begin{equation}\label{eq:2.2.11}
\begin{split}
E(y) &\approx \overline{y} +\frac{\sigma_y^2\rho'}{\sqrt{2\pi}} \int\limits_{-\infty}^{\infty} z^2 e^{-z^2} dz\\
&= \overline{y} +\frac{\sigma_y^2\rho'}{\sqrt{2\pi}} \int\limits_{-\infty}^{\infty} e^{-z^2} dz\\
&= \overline{y} +\sigma_y^2\rho'.
\end{split}
\end{equation}

Density bias (section 2.9) arises because the stellar distribution is not uniform, but is concentrated in the Galactic plane. Other variations in stellar density due, for example, to spiral structure, will also lead to a bias, but greater knowledge of the stellar distribution is required for a correct treatment. 

For the sake of simplicity I will treat separately the different causes of biases which affect the estimation of apparent magnitude and distance of stars. The corrections calculated here can be regarded as first order corrections. As these first order corrections are generally less than 10\%, and typically less than 5\%, it is reasonable to ignore and higher order terms in equation (\ref{eq:2.2.6}) and due to the non-linearity of stellar density, $\rho$.

\subsection{Parallax bias}\label{sec:2.3}
It is well known that the inverse relationship between distance and the measured quantity, parallax, leads to a bias in stellar distances. Suppose that a star's actual distance is $R$ pc, its true parallax is $ p = 1000 / R $ and that the result of a measurement of $ p $ is $ \pi $, where $ \pi $ is normally distributed with mean $ p $ and standard deviation $ \sigma $. The probability density for $ \pi $ is $ f(\pi) = N(\pi, p, \sigma) $. Restricting to positive parallax, expected measured distance is:
\begin{equation}\label{eq:2.3.1}
\begin{split}
E(\frac{1000}{\pi}) &= \int\limits_{0}^{\infty} \frac{1000}{\pi} N(\pi,p,\sigma) d\pi \\
&= \frac{1000}{p} \int\limits_{0}^{\infty} \frac{1}{x} N(x,1,\frac{\sigma}{p}).
\end{split}
\end{equation}
\begin{figure}
	\centering
		\includegraphics[width=0.47\textwidth]{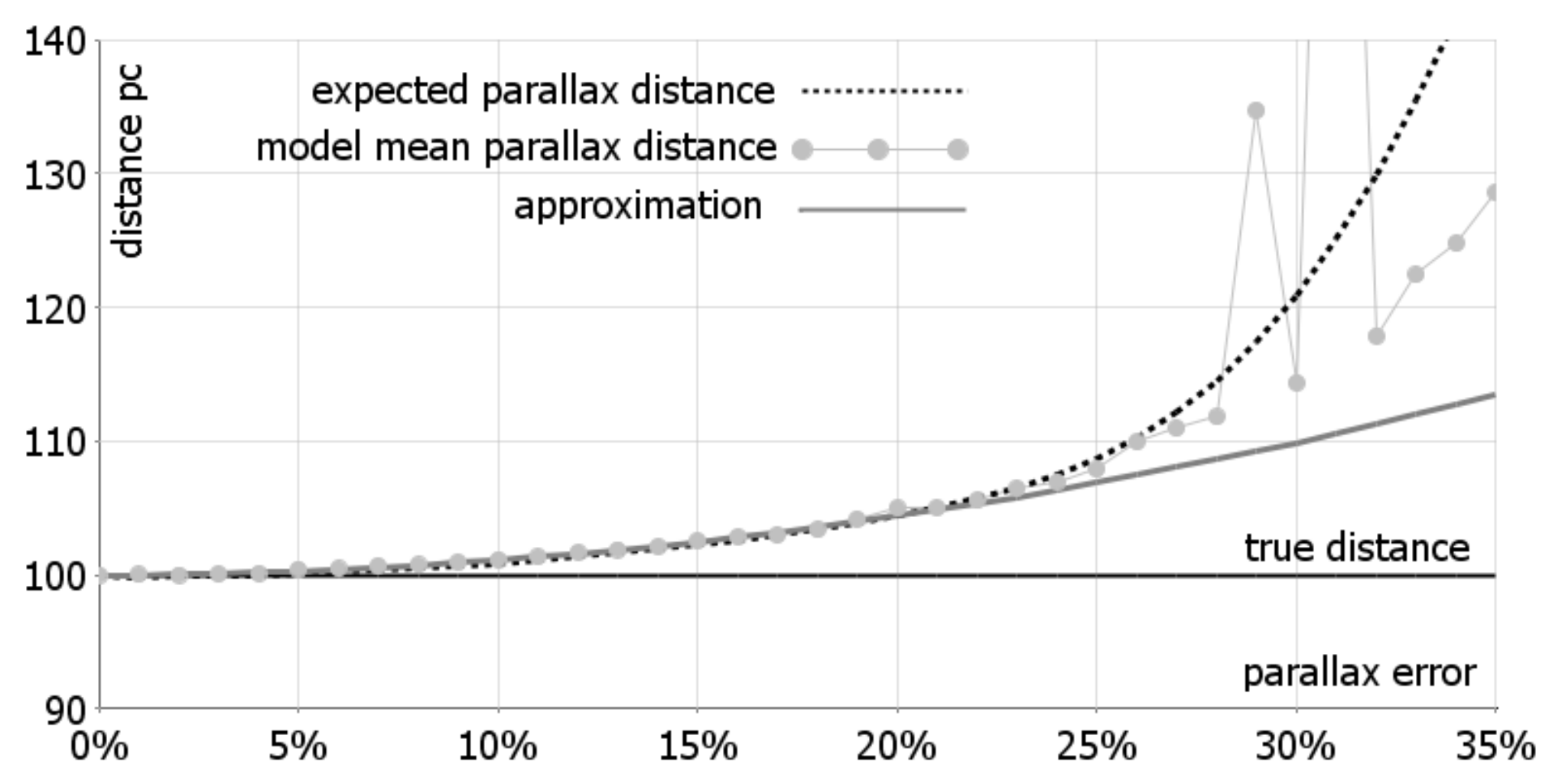}
	\caption{Expected parallax distance for a star at true distance 100 pc (dashed black, equation (\ref{eq:2.3.1})) compared to the quadratic approximation (continuous grey, equation (\ref{eq:2.3.2})), and the mean distance for numerical distributions of 10 000 stars at 100 pc with random Gaussian errors in parallax (light grey dots). Above about 27\% parallax error, the mean parallax distance for the model is very uncertain. }
	\label{Fig:1}
\end{figure}
\begin{figure}
	\centering
		\includegraphics[width=0.47\textwidth]{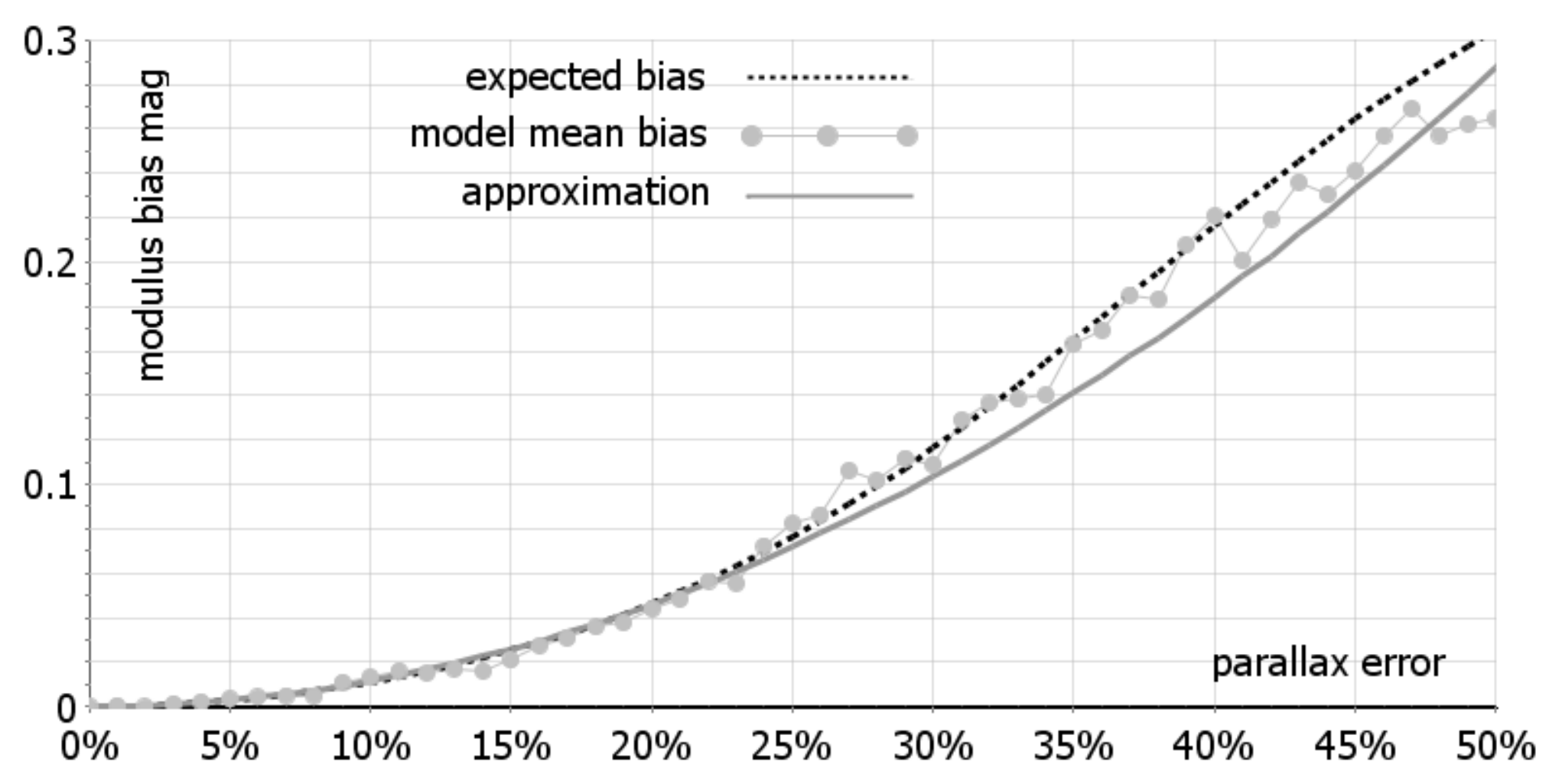}
	\caption{Expected modulus bias (dashed black, integral in equation (\ref{eq:2.4.1})) compared to a quadratic approximation (continuous grey, equation (\ref{eq:2.4.2})), and the mean modulus bias for numerical distributions of 10 000 stars with random Gaussian errors in parallax (light grey dots). }
	\label{Fig:2}
\end{figure}
The integral in equation (\ref{eq:2.3.1}) can be evaluated numerically (figure 1). For $ \sigma /p <\; \sim 25 $\%, 
\begin{equation}\label{eq:2.3.2}
E(\frac{1000}{\pi}) \approx R(1+1.1\frac{\sigma^2}{p^2}).
\end{equation}
Thus, measured distance is expected to overstate true distance. The condition, $ \sigma /p <\; \sim$25\%, is satisfied when $ \sigma / \pi < \;\sim$20\%. Then, since replacing $ \sigma / p $ by $ \sigma / \pi $ in equation (\ref{eq:2.3.2}) introduces only higher order terms, a corrected distance estimate for parallaxes with errors $ \sigma / \pi <\; \sim$20\% is:
\begin{equation}\label{eq:2.3.3}
R \approx \frac{1000}{\pi}\div(1+1.1\frac{\sigma^2}{\pi^2}).
\end{equation}

I tested this formula using a numerical model containing 10\,000 stars at 100 pc with a fixed parallax error. For each value of parallax error, a random Gaussian error was applied to the parallax for each star, and the mean distance was plotted for different values of the standard error. The model confirmed the validity of equation (\ref{eq:2.3.3}) up to about 25\% parallax errors, and showed that any distance estimate becomes highly unreliable when parallax errors approach or exceed 30\%. This is because the range of parallaxes within error bounds contains very small and negative values; $ R $ is a discontinuous function of $ \pi $ at $ \pi = 0 $. Calculation is thus invalidated. It remains reasonable to use parallax measurements to place a lower bound on distance, but in the absence of an upper bound, one should not give an estimate for a mean of the distance probability distribution.

\subsection{Modulus bias}\label{sec:2.4}
Similarly, the expected distance modulus is:
\begin{equation}\label{eq:2.4.1}
\begin{split}
E(\mu) &= \int\limits_{0}^{\infty}(5\log\frac{1000}{\pi}-5) N(\pi,p,\sigma) d\pi \\
&= \int\limits_{0}^{\infty}(5\log\frac{1000}{px}-5) N(x,1,\frac{\sigma}{p}) dx \\
&= \int\limits_{0}^{\infty}(5\log\frac{1000}{p}-5-5\log x) N(x,1,\frac{\sigma}{p}) dx \\
&\approx \mu_p - 5\int\limits_{0}^{\infty} N(x,1,\frac{\sigma}{p}) \log x dx
\end{split}
\end{equation}
where $ \mu_p $ is the distance modulus for a star with true parallax $ p $, and noting that the integral of the normal distribution approaches unity when $ p \gg \sigma $. Thus, modulus bias is given by the integral in equation (\ref{eq:2.4.1}), which can be evaluated numerically (figure 2). A quadratic approximation is given by
\begin{equation}\label{eq:2.4.2}
\mathrm{modulus\;bias} \approx 1.15\frac{\sigma^2}{\pi^2}.
\end{equation}
I tested this formula using a numerical model containing $ 10\,000 $ stars at 100 pc, with Gaussian parallax. I plotted mean distance modulus error against standard error. This confirmed equation (\ref{eq:2.4.2}) to above $ 25 $\% parallax errors. Above parallax errors of $ \sim$40\%, the effect of truncation of the normal distribution in equation (\ref{eq:2.4.1}) becomes apparent.

\subsection{Magnitude bias}\label{sec:2.5}
In practice, quoted errors in magnitudes in 2MASS are sufficiently small that they generate distance errors in the order of one percent. Any resulting bias will be an order of magnitude less, and can be ignored. Much greater errors arise from the real differences in the magnitudes of stars of given colour due to such causes as age and composition. The magnitude distribution (seen, for example, in figure 5) is clearly not symmetrical across the main sequence, and in addition the relationship between distance and magnitude follows a non-linear law (actually exponential). If, for a given colour and stellar class, the distribution, $ u(M) $, of absolute magnitudes, $ M $, has mean $ \overline{M} $ and is contained in the interval $ [M_{\mathrm{min}} , M_{\mathrm{max}}] $ then the expected distance for a star of visual magnitude $ m $ is 
\begin{equation}\label{eq:2.5.1}
\begin{split}
E(R) &= \int\limits_{M_{\mathrm{min}}}^{M_{\mathrm{max}}} 10^{(m-M+5)/5} u(M)dM \\
 &= 10^{(m-\overline{M} + 5) / 5} \int\limits_{M_{\mathrm{min}}}^{M_{\mathrm{max}}} 10^{(\overline{M}-M)/5}u(M) dM
\end{split}
\end{equation}

Thus, there is a bias in luminosity distance due to physical differences in stars. The magnitude bias factor, $l$, is given by the integral in equation (\ref{eq:2.5.1}), and depends on the distribution of magnitudes within a stellar class, as well as on magnitude measurement errors (which are small compared to physical differences). It is not possible to evaluate this integral without prior knowledge of $u$. The calibration here removes magnitude bias empirically, by minimising the sum of squared differences between luminosity and parallax distances for the calibration sample.

\subsection{Malmquist bias}\label{sec:2.6}
The Malmquist bias (1922, 1925) arises because brighter stars at given distance are more visible than less bright stars. The mean absolute magnitude of a magnitude limited sample of any given stellar class is less than that of a distance limited sample because the magnitude limited sample includes a greater proportion of brighter stars at greater distances. The effect of the Malmquist bias will be seen in figure 5, in which the best fit regression for a magnitude limited sample lies above the mean position of stars with better than 3\% parallax errors. Based on approximate assumptions, Malmquist calculated that the bias in magnitude should be $ \overline{\Delta M} = -1.382\sigma^2$. The Malmquist correction must be applied when the calibration sample is distance limited. For Hipparcos red giants with 10\% parallax errors, this gives $ \overline{\Delta M} \approx -0.20 $ mag, and for blue dwarfs with 3\% parallax errors, $ \overline{\Delta M} \approx -0.23 $ mag, but these figures depend on assuming uniform stellar density and a Gaussian luminosity function.

Consider stars of particular colour and class. The distribution function for the absolute magnitude of these stars is defined on some interval. Then the distribution function of distances of stars of this class with given visual magnitude, $ m_1 $, is also defined on an interval, and the expected distance calculated for a star of this class will take into account that the differential volume element, $ dV = r^2drd\Omega $ for solid angle $ d\Omega $, increases with $ r^2 $. The volume in space described by a given solid angle and visual magnitude $ m_2 $ is geometrically similar, so that mean distance is greater by a factor $ 10^{(m_2-m_1)/5} $. Thus, if the calibration sample and target population are both magnitude limited, the Malmquist bias does not appear.

Hipparcos is complete to $ V =\; \sim$7.3 - 9 mag, depending on Galactic latitude and spectral type (Perryman et al. 1997). If the calibration sample is limited by parallax accuracy, for example to 5\% parallax errors as done by Bilir et al. (2008), then the sample will be biased toward low distances (it would be distance limited if all parallax errors were identical, but this is only approximately true). In practice, because of the approximations in the calculation of the correction, because magnitude depends on stellar type, because parallax errors are not uniform, because about half of the Hipparcos catalogue is an incomplete sampling of objects to a limiting magnitude of about 12 mag, and because the distribution of stars in space is not uniform, it is not possible to estimate the Malmquist correction accurately.

As described in section 4.1 and section 4.2, the calibration sample is magnitude limited to ensure that distances are inside the limit for stars with 20\% parallax errors. This minimises the Malmquist bias and leaves a population of sufficient size that the error in mean magnitudes is small. For red giants, comparison with a population limited to 10\% parallax errors shows that the actual level of Malmquist bias is $ \overline{\Delta M} \approx -0.15 $ mag (a 10\% bound is probably not small enough to show the full Malmquist bias, but so few giants with smaller parallax errors have good magnitudes that a smaller bound is not meaningful). Comparison with a population of blue dwarfs limited to 3\% parallax errors finds $ \overline{\Delta M} \approx -0.63 $, more than twice the theoretical estimate and equivalent to a systematic error in distances of nearly 30\%. In part this arises from the non-Gaussian nature of the magnitude distribution for dwarfs, but the major reason appears to be the non-uniform distribution of star forming regions. Figures for both dwarfs and giants are subject to random variations depending on the precise parameters used to limit the population.

\subsection{Trumpler-Weaver bias}\label{sec:2.7}
The Trumpler-Weaver (1953) bias (here distinguished from the Lutz-Kelker bias, section 2.8) is not an error in individual stars but is a selection bias affecting the mean parallax distance of a population within a sphere of given radius. The number of stars with true distances greater than this radius which appear in the sample due to parallax error will exceed the number with true distances inside the sphere whose parallax errors remove them from the sample, because the volume of the error shell outside the sphere is greater than the volume of the error shell inside the sphere. The consequence is that the true mean distance of the sample is greater than mean parallax distance. 
\begin{figure}
	\centering
		\includegraphics[width=0.47\textwidth]{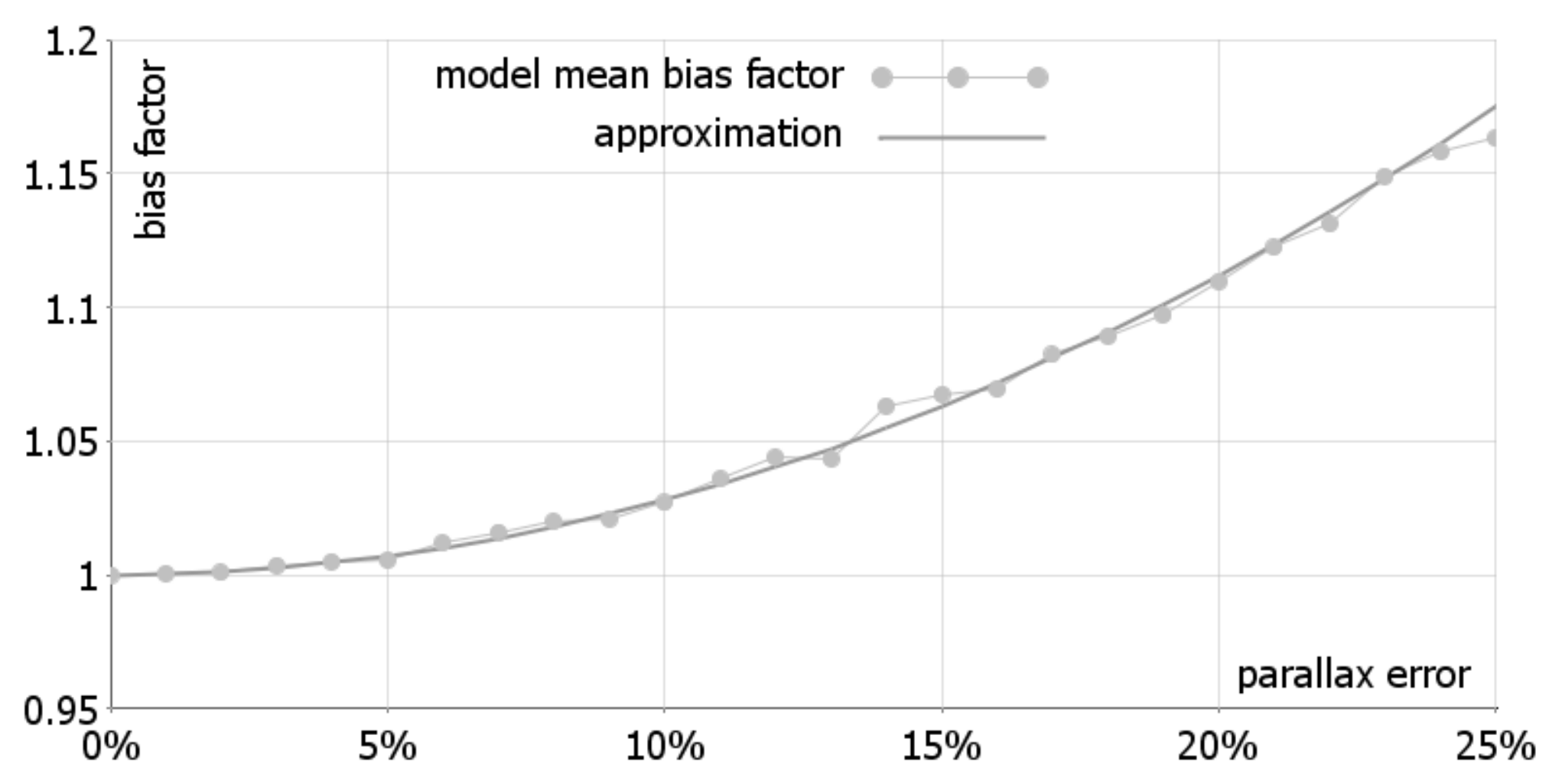}
	\caption{ The Trumpler-Weaver bias for a numerical model with a uniform stellar distribution (dots), and the quadratic approximation given by equation (\ref{eq:2.7.1}) (continuous).}
	\label{Fig:3}
\end{figure}

To determine the effect of the Trumpler-Weaver bias I created a numerical model with a uniform distribution of 30\,000 stars spread over a cube of side 400 pc (large enough that the boundaries of the cube will not affect the result). A star with true distance $R$ pc, has true parallax $ p = 1000 / R $. I generated nominal measured parallaxes, $ \pi $, by adding a random error to $ \pi $ for each star, using a Gaussian distribution with standard error $ 0.01np $ for integral values of $ n $. I then truncated the distribution using $ \pi > 10 $ and plotted the ratio of true distances to parallax distances, i.e. $ \pi / p $, for the stars in the truncated distribution against $ n $\% (figure 3). I fitted a quadratic approximation to the resultant curve, 
\begin{equation}\label{eq:2.7.1}
\pi/p \approx 1 + 0.00028n^2.
\end{equation}
Thus, the Trumpler-Weaver bias is estimated at 0.25\% for parallax errors of 3\%, and 2.8\% for parallax errors of 10\%. The formula is independent of cutoff distance.

\subsection{Lutz-Kelker bias}\label{sec:2.8}
Lutz and Kelker (1973, 1974, 1975) observed that, when a distance limited sample has been used for calibration, the Trumpler-Weaver bias will generate a systematic error in the luminosity distances of individual stars, and calculated the appropriate correction to magnitude (note that Binney and Merrifield 1998 do not correctly describe the bias discussed by Lutz \& Kelker. See also Haywood Smith 2003). The bias of equation (\ref{eq:2.7.1}) is removed by adding a correction $ \Delta M $ to absolute magnitude where
\begin{equation}\label{eq:2.8.1}
\Delta M = -5.8 (\overline{\sigma / \pi})^2.
\end{equation}
Although the Lutz-Kelker bias is small, perhaps negligible, for a population with small parallax errors, the restriction to small parallax errors makes it impossible to avoid the Malmquist bias. For an accurate calibration of luminosity distances, it is necessary use a magnitude limited calibration sample. 

\subsection{Density bias}\label{sec:2.9}
It is more probable to find a star in a region of high stellar density than in one of low density. Thus, a star found near a position of high density is likely to be nearer to that position than measurement of its heliocentric distance suggests. In practice, only an approximation is possible because stellar density varies with spiral structure which is not well known, and contains random as well as systematic variations. For definiteness, I adopt a density model for the solar neighbourhood containing thin and thick disc components such that the variation in density with height $ z $ above the Galactic plane is
\begin{equation}\label{eq:2.9.1}
\rho(z) = \frac{1}{20}(19 e^{-|z|/300} + e^{-|z|/1500})
\end{equation}
\begin{equation}\label{eq:2.9.2}
\rho'(z) = \frac{-\mathrm{sgn}z}{30000}(95 e^{-|z|/300} + e^{-|z|/1500})
\end{equation}
The velocity gradient radially to the Sun is $ \rho'\sin b $ where $ b $ is Galactic latitude.

Although the error distribution for stellar distance is not normal, it may be approximated by a normal distribution in a first order estimation of the density correction. Suppose a star is found at position $R$ and that the fractional error $ \epsilon $ is normally distributed with pdf $ N(\epsilon, 0, \sigma) $. Suppose too that stellar density in the radial direction is, for small fractional displacement $ \epsilon $,
\begin{equation}\label{eq:2.9.3}
\rho(R+\epsilon R) \approx \rho_0 + \epsilon R\rho' \sin b 
\end{equation}
where $ \rho' $ can be taken as constant. The dependency of the differential volume element on $ r^2 $ has already been considered in the treatment of the Malmquist bias and is not required again here. Equation (\ref{eq:2.9.3}) is essentially the same as equation (\ref{eq:2.2.11}) up to normalisation. Thus
\begin{equation}\label{eq:2.9.4}
E(R) \approx R(1+\frac{\sigma^2 R \rho' \sin b}{\rho_0})
\end{equation}

I applied this formula to both parallax distances and luminosity distances in the calibration sample. For the majority of stars the correction is less than half of one percent of distance, but for a few stars at high latitudes with large parallax errors the correction is above 5\%. The correction altered the overall error estimate by 0.2\%. I also applied the correction to RAVE luminosity distances. The greatest corrections are for blue dwarfs, for which 5\% is typical and 15\% is maximal. For other classes, 2\% corrections are typical and 4\% maximal.

\subsection{Aging bias} \label{sec:2.10}
In the main sequence, older stars are brighter than younger stars with the same mass and composition, but in the absence of precise age estimates for individual stars it is impossible to take account of the effect of age on the estimate of absolute magnitude. Both RAVE and Hipparcos contain predominantly thin disc stars. Differences in stellar velocity ensure mixing of stars of all ages and compositions in the thin disc in all sky regions (except for stars in clusters). The effect of age on magnitude is already taken into account within the calculated errors for the calibration sample, and there is no basis on which we may claim a systematic difference in ages for the bulk of thin disc stars.

Because RAVE scanned a region of sky away from the Galactic plane, and because RAVE stars are more distant, the proportion of thick disc stars is greater in RAVE. This is compensated by the selection of halo stars for Hipparcos. Halo and thick disc stars are a small proportion of both populations, and will not have a great effect on mean luminosity. It is also true that stars in fast moving streams are older than the bulk of the population. However, the proper identification of streams requires prior knowledge of kinematic properties, and particularly knowledge of distance. For this reason, in the prepared data base no age correction is used. Researchers studying the thick disc and fast moving streams may apply their own magnitude correction to take account of the mean age of these stars. 
\begin{figure}
	\centering
		\includegraphics[width=0.47\textwidth]{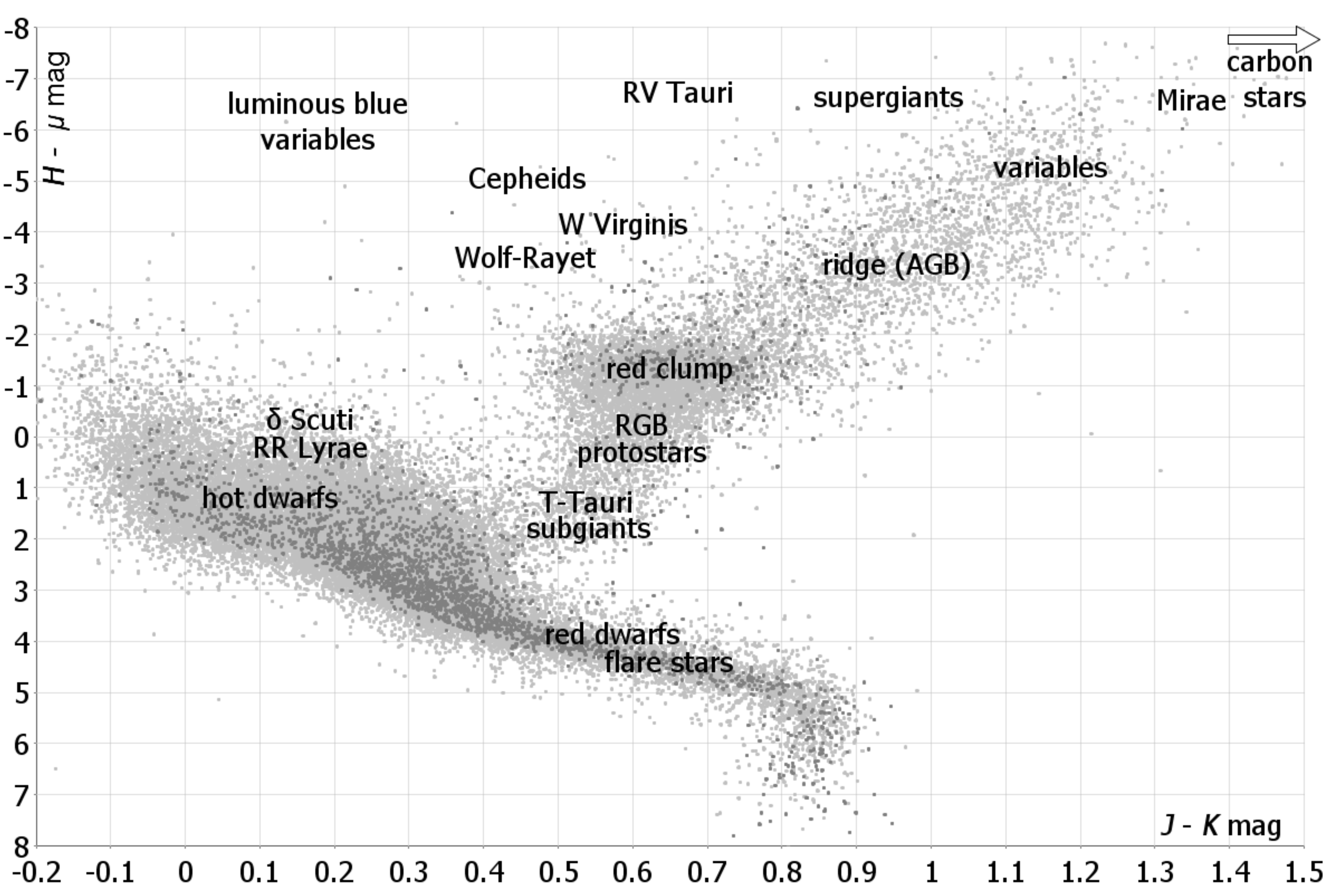}
		\includegraphics[width=0.47\textwidth]{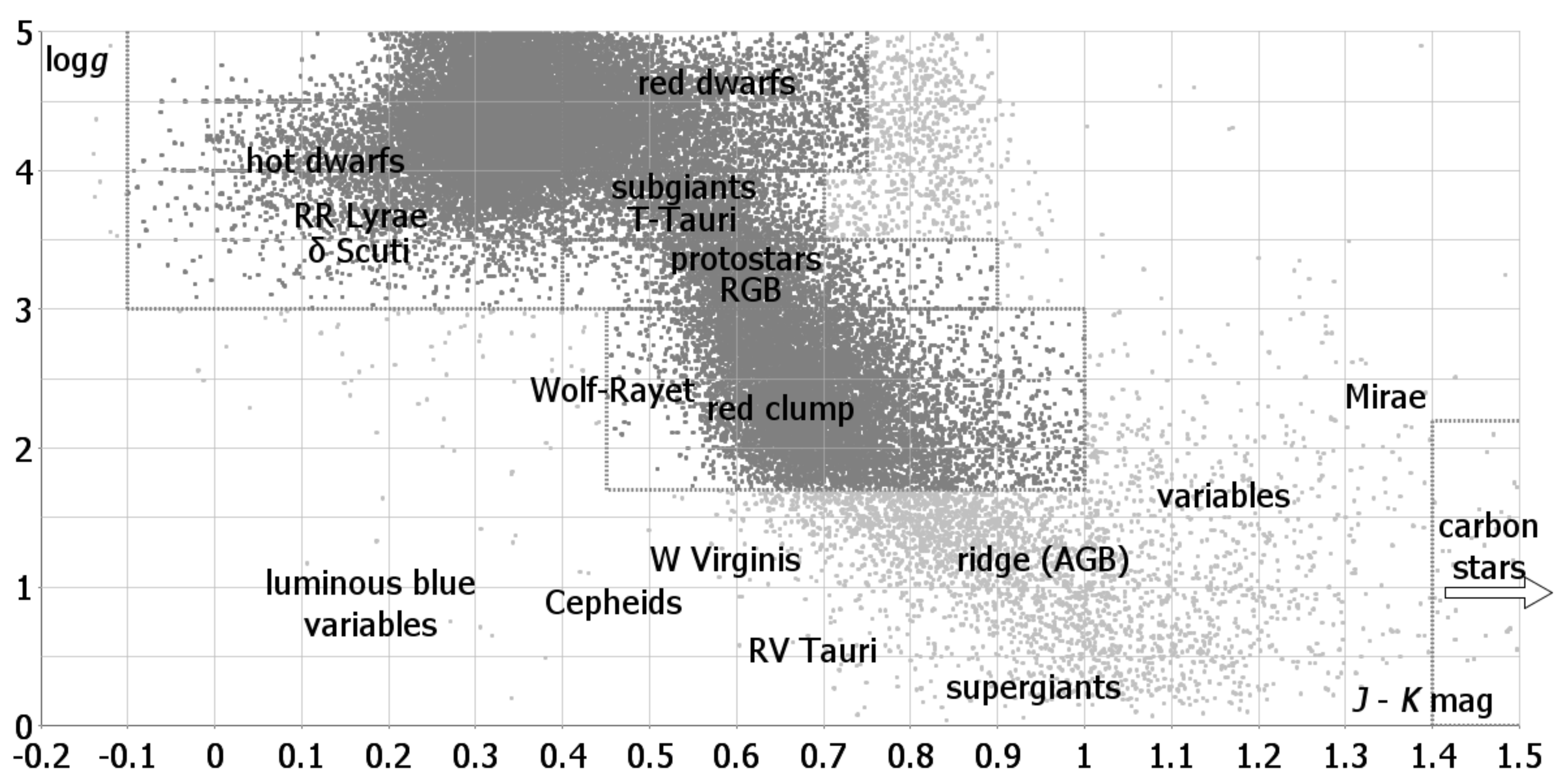}
		\caption{Top: stellar classes shown on a CMD of Hipparcos stars with absolute $ H $ magnitude against $ J - K $. Stars with better than 3\% parallax errors are shown in dark grey and those with better than $ 15\% $ errors in light grey. Dereddening has been applied to stars beyond 100 pc. Approximate positions of stellar classes have been checked by reference to CDS. The position of particular classes with respect to the major features depends on the choice of filters (in particular, Mirae and Wolf-Rayet stars are substantially displaced on a diagram for absolute $ Hp $ magnitude against $ B - V $). Bottom: classes shown on a plot of $ \log g $ against $ J - K $ for RAVE stars. Upper and lower plots use the same scale on the colour axis to emphasise the correspondence between the main populations. The dotted lines show boundaries for class regions used to calculate distances for stars shown in dark grey.}
	\label{Fig:4}
\end{figure}

\section{Stellar classification}\label{sec:3}
Stellar evolutionary theory has made it possible to identify stars at different stages of stellar evolution according to position on a CMD. In figure 4 (top) the positions of particular classes have been checked by reference to CDS. The plot of $\log g$ against $J-K$ for RAVE stars (figure 4, bottom) similarly shows well defined populations. The same scale on the colour axis is used for both plots, and it is seen that in a rough sense the one plot is a reflection of the other in a horizontal axis. Indeed, since the diameter of a star is directly related to the heat which it generates, we expect a negative correlation between absolute luminosity and surface gravity. There is thus a direct correspondence between the Hipparcos population on the CMD in the upper plot and the RAVE population on the colour-gravity diagram in the lower plot, and we can identify evolutionary classes in RAVE according to $\log g$ and $J-K$ by comparing the plots. We can therefore estimate absolute magnitude for RAVE stars and hence calculate stellar distance.

Breddels et al. (2010), Zwitter et al. (2010), and Burnett et al. (2011) used a range of parameters, including metallicity and surface gravity to identify stellar class via isochrones. However, as seen in figure 4, surface gravity alone gives a clear demarcation into classes, whereas metallicity takes a range of values in all parts of the CMD, and was found to have very little effect on distances using $H$ magnitudes, $ M_{H} $, and $J-K$ colours (section 4.1). For this reason I believe that using only $\log g$ with $J-K$ colours gives more accurate identification of stellar class.
\begin{table}\label{(table1)}
\begin{scriptsize}
\begin{flushleft}
\begin{tabular}{lrrrrr} 
\textbf{Class} & $ J-K_{\mathrm{min}} $ & $ J-K_{\mathrm{max}}$ & $ \log g_{\mathrm{min}} $ & $ \log g_{\mathrm{max}x} $ & count\\
hot dwarf & $ -0.1 $ & 0.4 & 3 & 5 & 16 395\\
red dwarf & 0.4 & 0.75 & 4 & 5 & 5 487\\
TT\&S & 0.4 & 0.7 & 3.5 & 4 & 1 969\\
PS\&R & 0.4 & 0.9 & 3 & 3.5 & 2 263\\
RC & 0.45 & 1.0 & 1.7 & 3 & 9 054\\
\end{tabular}
\caption{Bounds in colour and surface gravity for five stellar classes represented in RAVE. The count is the number members found in RAVE after calculation of distance and dereddening.}
\end{flushleft}
\end{scriptsize}
\end{table}

I restricted the RAVE data base as described in section 5.1, and identified five populations of RAVE stars (figure 4, bottom, and table 1) for which we can give reasonably precise estimates of absolute magnitude, hot dwarfs, red dwarfs, T-Tauri and subgiants, protostars and red giant branch, and the red clump. It is also possible to identify asymptotic giant branch stars, but, due to variability and the range of stellar processes found in the AGB, absolute magnitudes cover a broad range of values and the resulting estimate of distance shows more than 40\% errors. There is some indication in Hipparcos that carbon stars have sufficiently uniform absolute magnitudes to have some value as standard candles, but Hipparcos AGB stars (including carbon stars) are so bright that 2MASS $J$, $H$ and $K$ magnitudes may be subject to systematic error due to saturation of detectors. I have therefore not calibrated luminosity distances for about 50 carbon stars which appear in RAVE. I have calibrated distances for white dwarfs, but they do not appear to be represented in RAVE (being insufficiently luminous).

It will be seen that the velocity distributions for protostars and red giant branch and for T-Tauri and subgiants have a substantial peak close to circular motion, showing the kinematics of new stars (section 5.2). The characterisation given here in terms of position on the CMD is for the purpose of determining luminosity distances and does not constitute a rigorous determination of evolutionary phase. Most sources do not define the term `protostar' with precise evolutionary endpoints, but typically the end of the Hayashi track and the beginning of the Henyey track is taken as the endpoint of protostar evolution. Prialnik (2010) specifically describes the Hayashi track as the protostar phase, and terms the preceding phase `stellar embryo', while other authors (e.g., Kippenhahn and Weigart 1990) use protostar to describe both phases. The terminology used here reflects the fact that the colour and magnitude of these groups correspond closely to the theoretical effective temperature and luminosity of Hayashi tracks (protostars) and Henyey tracks (T-Tauri).

\section{Distance calibration}\label{sec:4}
\subsection{Calibration for dwarfs}\label{sec:4.1}
For stars within about 100 pc (the local bubble) dereddening is not required. I dereddened the remaining Hipparcos population for Galactic latitudes greater than $ b = 9.7 $\textdegree using the maps of Burstein \& Heiles (1978, 1982), together with Bahcall \& Soneira's (1980) formula,
\begin{equation}\label{eq:4.1.1}
A_d(b)=A_{\infty}(b)(1-\exp(\frac{-|d\sin b|}{h})),
\end{equation}
where $ A_{\infty}(b) $ and $ A_d(b) $ are total absorption at infinity and at stellar distance $ d $; $ A_{\infty}(b) =3.1E_{\infty}(B-V)$ is found from the reddening map; $ h = 125 $ pc is the adopted scale height for interstellar dust (Marshall et al. 2006). Absorption in each magnitude is found using factors given by Schlegel et al., $ A_J =0.902 E(B - V) $, $ A_H = 0.576 E(B - V) $, $ A_K = 0.367 E(B - V) $. Burstein \& Heiles map is of lower resolution than that of Schlegel, Finkbeiner \& Davis (1998), but is compatible for Galactic latitudes above $ b = 9.7 $\textdegree and is empirically based on the reddening of other galaxies, rather than calculated theoretically from dust maps. As the effect of reddening on distances is not large outside of the Galactic plane, this choice is likely to have little practical impact.
\begin{figure}
	\centering
		\includegraphics[width=0.47\textwidth]{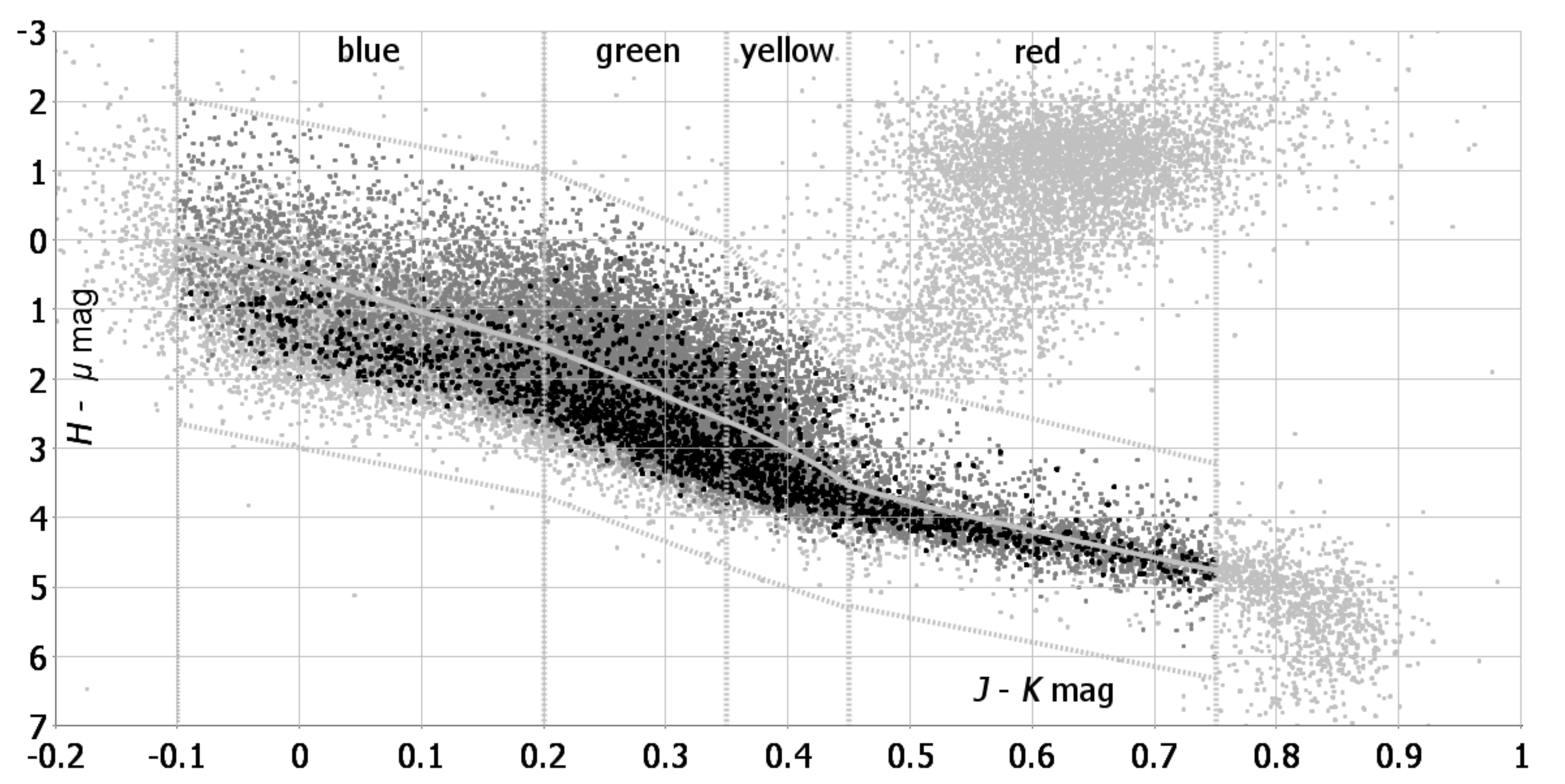}
		\caption{The CMD for absolute $ H $ magnitude against $ J - K $ mag for main sequence Hipparcos stars with parallax errors better than 20\% and 2MASS `AAA' quality rating. Stars used in the calibration (dark grey and black) are magnitude limited by equation (\ref{eq:4.1.5}). Pale greyed stars are excluded from calibration. Stars with better than 3\% parallax errors are shown in black. The bulk of main sequence stars are in the dense region toward the bottom of the band. It is not possible to avoid the inclusion of some T-Tauri and subgiants. The best fit regression is shown as a continuous grey line, in which quadratic correction terms have been used to ensure continuity at band boundaries. The Malmquist bias is seen because the best fit regression is above the mean position of stars with better than 3\% parallax errors.}
	\label{Fig:5}
\end{figure}
\begin{table}\label{(table2)}
\begin{small}
\begin{flushleft}
\begin{tabular}{lrrrr} 
\textbf{Band} & \textbf{red} & \textbf{yellow} & \textbf{green} & \textbf{blue}\\ 
lower colour bound, $ b_{\mathrm{lo}} $	 & 0.45	 & 0.35	 & 0.20	 & $ -0.10 $\\
upper colour bound, $ b_{\mathrm{hi}} $	 & 0.75	 & 0.45	 & 0.35	 & 0.20\\
centre, $ p = (b_{\mathrm{lo}} +b_{\mathrm{hi}})/2 $	 & 0.60	 & 0.40	 & 0.275	 & 0.05\\
 \\
$ m_{1} $	 	 & $ 3.5 $	 & $ 6.0 $	 & 6.5	 & 3.5\\
$ c_{1} $	 	 & 3.7	 & 2.6	 & 2.4	 & 3.0\\
$ m_{2} $	 	 & 4.3	 & 19.0	 & 7.0	 & 3.5\\
$ c_{2} $	 	 & 0.0	 & $ -6.6 $	 & $ -2.4 $	 & $ -1.7 $\\
\\
slope, $ m $	 & 3.85	 & 8.89	 & 7.46	 & 5.22\\
intercept, $ c $		 & 1.88	 & $ -0.54 $	 & 0.02	 & 0.52\\
$ a_{\mathrm{hi}} $	 	 & 0.00	 & 31.17	 & $ -5.15 $	 & $ -1.25 $\\
$ a_{\mathrm{lo}} $	 	 & $ -3.46 $	 & 11.59	 & 5.02	 & 0.00\\
\\
mag bias factor, $ l $	 	 & 0.95	 & 0.89	 & $ 0.88 $	 & 0.86\\
$ d_{\mathrm{hi}} $	 	 & 0.00	 & $ 12.73 $	 & 1.11	 & 0.35\\
$ d_{\mathrm{lo}} $	 	 & $ -1.41 $	 & $ -2.49 $	 &$ -1.41 $	 & $ 0.00 $\\
\\
number of stars	 & $ 2\,391 $	 & $ 3\,351 $	 & $ 7\,309 $	 & $ 3\,769 $\\
error	 	 & $ 9.9 $\%	 & $ 18.8 $\%	 & $ 23.1 $\%	 & $ 24. $7\%
\end{tabular}
\caption{Parameters for the calibration of luminosity distance to parallax distance for the four colour bands used for the main sequence. The meaning and use of the parameters is as described in the text. The error is for stars with better than 2\% parallax errors.}
\end{flushleft}
\end{small}
\end{table}

I split the population into four bands by colour, designated for convenience as red, yellow, green and blue, using 
\begin{equation}\label{eq:4.1.2}
b_{\mathrm{lo}} < J - K < b_{\mathrm{hi}}.
\end{equation}
where $ b_{\mathrm{hi}} $ and $ b_{\mathrm{lo}} $ are given in table 2. To restrict the population to dwarfs, I placed bounds above and below the main sequence (figure 5),
\begin{equation}\label{eq:4.1.3}
m_1(J - K) + c_1 > H - \mu > m_2(J - K) + c_2.
\end{equation}
\begin{figure}
	\centering
		\includegraphics[width=0.47\textwidth]{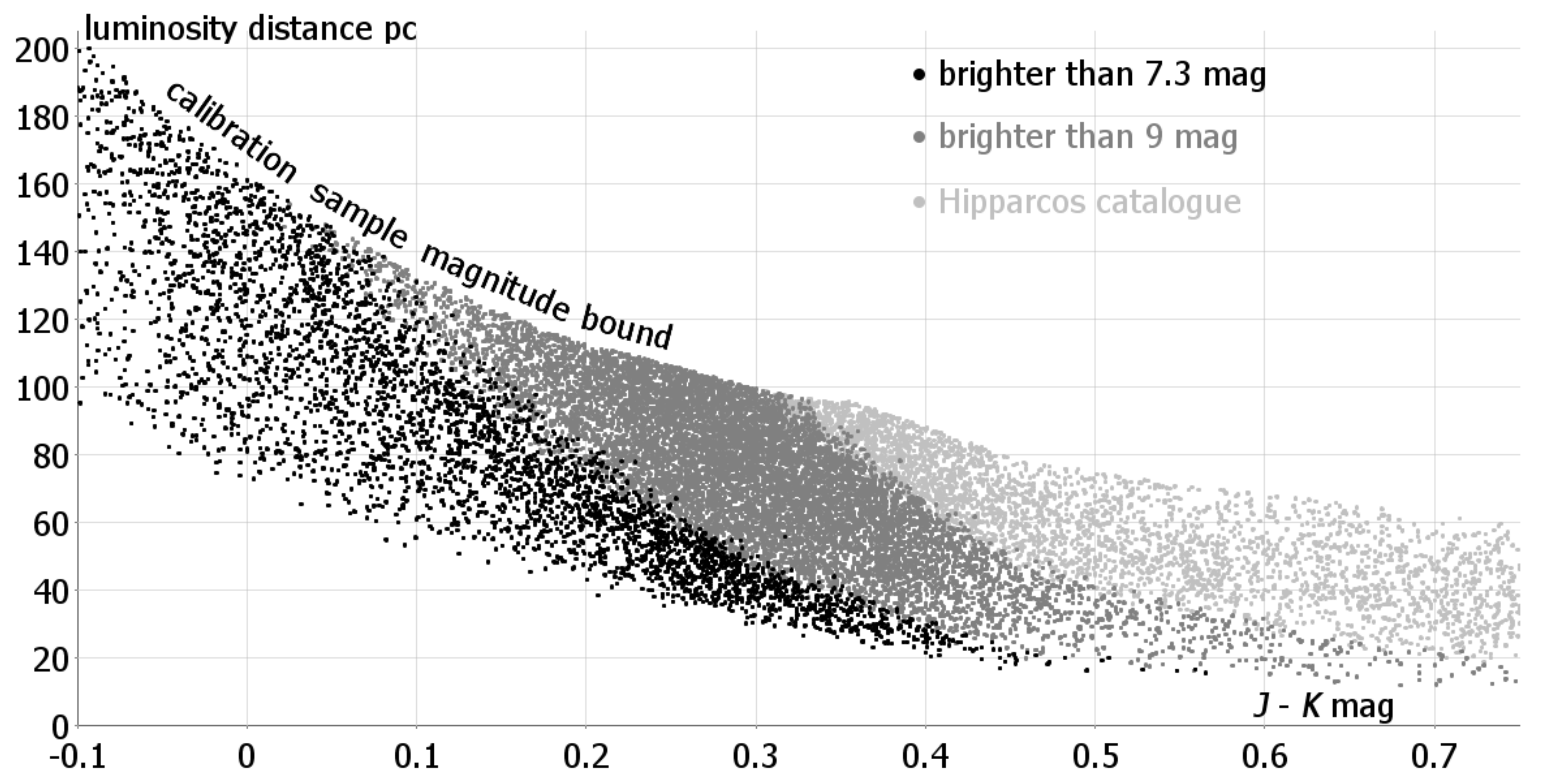}
		\caption{luminosity distance plotted against $ J - K $ mag for Hipparcos dwarves with parallax errors up to 20\%, and with magnitude cut-offs at $ Hp = 7.3 $ mag (black dots) and at $ Hp = 9 $ mag (mid grey dots). }
	\label{Fig:6}
\end{figure}
It is necessary to determine a bound such that the sample at each colour is magnitude limited, rather than limited by distance or by parallax error. Parallax errors for Hipparcos dwarfs are strongly correlated with colour, as brighter stars have more accurate parallaxes. Simple linear regression on the data gives the relation
\begin{equation}\label{eq:4.1.4}
\overline{ePlx} \approx 0.88(J-K) + 0.65.
\end{equation}
At all colours, the standard deviation of parallax error is close to one third of parallax error, so doubling the expected parallax error, $ \overline{ePlx} $, and multiplying by five gives a lower bound on parallax, above which almost all stars will have parallax errors of better than 20\%. I estimated the corresponding $H$ magnitude,
\begin{equation}\label{eq:4.1.5}
H_{\mathrm{max}} \approx m_1(J-K) +c_1 +5\log (\frac{100}{\overline{ePlx}}) -7,
\end{equation}
and used this to magnitude limit the calibration sample (figure 6). Equation (\ref{eq:4.1.5}) uses an approximation to the best fit regression line (which cannot be used because of circularity). This is adequate for estimating the magnitude bound, $ H_{\mathrm{max}} $, such that almost all stars brighter than $ H_{\mathrm{max}} $ have parallax errors better than 20\%. It is seen that $ H_{\mathrm{max}} $ is within Hipparcos' upper completeness bound of 9 magnitudes for colours $ J - K <\;\sim$0.33 mag. Since the Hipparcos population for redder stars is not distance limited it is reasonable to assume that it will not introduce Malmquist or similar bias. The incomplete sampling of Hipparcos stars with magnitudes above the completeness bound is chosen to ensure a representative population of stars of all types, including red and white dwarfs which would otherwise be excluded. With the exception of choosing a number of high proper motion stars (this is a small number compared to sample size), bias is likely to be towards choosing brighter representatives, i.e. toward magnitude limiting rather than distance limiting.

I used simple linear regression to find the line of best fit for absolute magnitude within each colour band, described by slope, $ m $, and intercept, $ c $. To ensure continuity at band boundaries I split the difference and added a quadratic term, obtaining for the curve of best fit,
\begin{equation}\label{eq:4.1.6}
M_{H} = H - \mu = m(J - K) + c + a(J - K - p)^2,
\end{equation}
where $ a = a_{\mathrm{hi}} $ when $ J - K > p $ and $ a = a_{\mathrm{lo}} $ when $ J - K < p $. This is shown by the continuous grey line in figure 5 (the quadratic term contributes little to distance estimates and does not affect calculated errors, but it removes ambiguity when corrections for reddening cause a change in colour band).

I used equation (\ref{eq:4.1.6}) to estimate absolute magnitude from colour, $J-K$, and hence calculated the distance modulus,$ \mu = H - M_{H} $, obtaining a first estimate of luminosity distance for each of the sample stars using,
\begin{equation}\label{eq:4.1.7}
R = 10^{\frac{\mu}{5}+1}.
\end{equation}
I corrected this estimate by using the magnitude bias factor, $l$, chosen to minimise the mean squared difference between luminosity distance and parallax distance for the calibration sample in each colour band. To remove magnitude bias, I multiplied distances given by equation (\ref{eq:4.1.7}) by $ l + d(J - K - p)^{2} $ with $ d = d_{\mathrm{hi}} $ when $ J - K > p $ and $ d = d_{\mathrm{lo}} $ when $ J - K < p $, where $ d_{\mathrm{hi}} $ and $ d_{\mathrm{lo}} $ are chosen to ensure the continuity of the correction factor at the band boundaries.

I tested calibrations using all combinations of the $J$, $H$, and $K$ magnitudes, as well as $ B - V $ and $Hpmag$. $ B - V $ gives slightly more accurate distances, but is not provided by 2MASS. $H$ outperformed other magnitudes, including $Hpmag$. The most accurate available colour is $J-K$. I tried improving the fit using a linear term for metallicity. This substantially reduces sample size and for $J-K$ colour it gives negligible improvement in accuracy. I therefore made no metallicity correction. Higher metallicity causes a star to be both redder and less luminous. If this shifts a star parallel to the main sequence on the CMD it will not affect the distance estimate. There is a small improvement in distances by correcting for metallicity when using the $ B - V $, but not when using $J-K$. 
\begin{figure}
	\centering
		\includegraphics[width=0.47\textwidth]{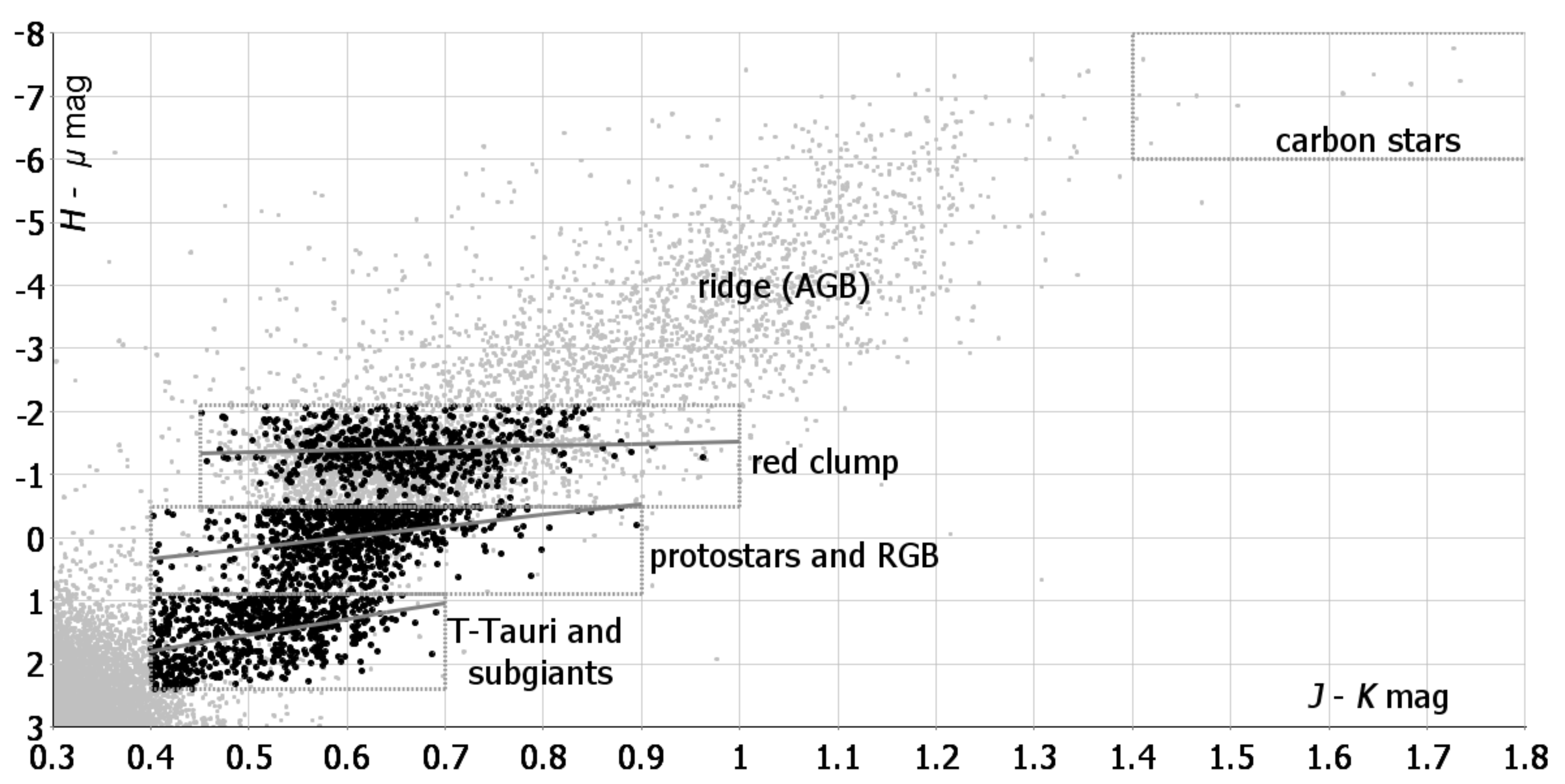}
		\caption{CMD for absolute $ H $ magnitude against $ J - K $ mag for Hipparcos giants and pre-main sequence stars with parallax errors better than 20\%. Dereddening has been applied for stars beyond 100pc. Grey dots represent stars with at least 2MASS `DDD' quality rating (valid JHK magnitudes). Black dots represent stars with 2MASS `AAA' quality rating used in the calibration. The best fit regression for three classes is shown by continuous grey lines.}
	\label{Fig:7}
\end{figure}
\begin{table}\label{(table3)}
\begin{flushleft}
\begin{tabular}{lrrrr} 
 & \textbf{TT\&S} & \textbf{PS\&R} & \textbf{RC} & \textbf{WD}\\ 
low bound, $ b_{\mathrm{lo}} $ 	 & $ 0.40 $	 & 0.40	 & 0.45	 & $ -0.25 $\\
high bound, $ b_{\mathrm{hi}} $ 	 & $ 0.70	 $ & 0.90	 & 1.00	 & 0.3\\
$ H_{\mathrm{max}} $ 	 	 &$ 7.18 $	 & 6.14	 & 5.09	 & 14.8\\
\\
$ c_{1} $	 	 & 2.40	 & 0.9	 & $ -0.50 $ & 14\\
$ c_{2} $	 	 & 0.9	 & $ -0.50 $	 & $ -2.10 $	 & 11\\
\\
slope, $ m $	 & $ -2.51 $	 & $ -1.76 $	 & $ -0.34 $	 & 5.45\\
intercept, $c$	 & 2.80	 & 1.04	 & $ -1.19 $	 & 12.37\\
\\
mag bias factor, $ l$	 & 0.96	 & 0.97	 & 0.97	 & 0.97\\
\\
count	 	 & $ 646 $	 & $ 795 $	 & $ 573 $	 & $ 12 $\\
error	 	 & 13.6\%	 & 11.4\%	 & 10.1\%	 & 14.3\%\\
\end{tabular}
\caption{Parameters for the calibration of luminosity distance to parallax distance for the T-Tauri and subgiants (TT\&S), protostars \& red giant branch (PS\&R), red clump giants (RC), and white dwarfs (WD).}
\end{flushleft}
\end{table}

\subsection{Calibration for other classes}\label{sec:4.2}
After correction for absorption, as above, I identified four stellar classes according to position on the CMD, T-Tauri and subgiants (TT\&S), protostars and red giant branch (PS\&R), the red clump (RC) and white dwarfs (WD) (figure 7 and table 3). Each class has colour bounds $ b_{\mathrm{lo}} $ and $ b_{\mathrm{hi}} $ and bounds on absolute magnitude, $ c_1 $ and $ c_2 $. 

For each class I found a parallax bound, $ Plx_{\mathrm{min}} $, equal to mean parallax error plus three times the standard deviation of the parallax error for stars in the class, multiplied by five. Almost all stars with parallaxes greater than $ Plx_{\mathrm{min}} $ have parallax errors better than 20\%. I estimated the corresponding $H$ magnitude,
\begin{equation}\label{eq:4.2.1}
H_{\mathrm{max}} \approx \frac{c_1 +c_2}{2} + 5 \log(\frac{1000}{Plx_{\mathrm{min}}} -5),
\end{equation}
finding $ H_{\mathrm{max}} = 7.18 $ mag for T-Tauri and subgiants, $ H_{\mathrm{max}} = 6.14 $ mag for protostars and red giant branch, and $ H_{\mathrm{max}} = 5.09 $ mag for the red clump.

After applying these magnitude limits, I used regression to find lines of best fit for each class. I then used equation (\ref{eq:4.1.6}) (with $ a = 0 $) to estimate absolute magnitude, from which an initial distance estimate was obtained. Luminosity distance is the initial distance estimate multiplied by the magnitude bias factor, $l$, chosen, as for dwarfs, to minimise the mean least squared difference between luminosity distance and parallax distance in the calibration sample.

\section{Application to RAVE}\label{sec:5}
\subsection{RAVE distances}\label{sec:5.1}
The RAVE third data release contains 83\,072 radial velocity measurements. I discarded repeated measurements on the same star, preferentially keeping those with values for MASK and $\log g$. I discarded entries which did not have `A' quality flags for cross referencing to 2MASS, or which did not have a `AAA' 2MASS quality index. I did not use RAVE quality flags at this point because the method is generally applicable in 2MASS, provided we have some means to determine stellar class with reasonable confidence (as we can for hot dwarfs). 

I further restricted the data base to the stellar classes described above. I combined the two populations of dwarfs into a single population. I used equation (\ref{eq:4.1.6}) to find absolute magnitude and calculated the distance by using equation (\ref{eq:4.1.7}) and multiplying by the magnitude bias factor. I then applied the density correction of equation \eqref{eq:2.9.4}

For Galactic latitudes $ |b| > 9.7 $\textdegree, I corrected for reddening using the iterative procedure described by Co\c{s}kuno\u{g}lu et al. (2011), in which absorption is calculated using equation (\ref{eq:4.1.1}) for the given Galactic longitude, latitude and stellar distance. Each magnitude is then corrected for absorption. Distance is then recalculated for the corrected colour. The procedure is repeated until distance converges to the required accuracy. RAVE contains a small number of stars with Galactic latitudes $ |b| \le 9.7 $\textdegree. I have only given distances when the first estimate is less than 100 pc, because there is no reliable estimate of the correction due to reddening in the Galactic plane.
\begin{table}\label{(table4)}
\begin{flushleft}
\begin{tabular}{lrrrr} 
 & \textbf{Dwarfs} & \textbf{TT\&S} & \textbf{PS\&R} & \textbf{RC} \\ 
count	 	 & 39 514	 & 1 961	 & 2 262	 & 9 057\\
mean pc	 	 & 360	 & 515	 & 928	 & 1 391\\
max pc	 	 & 2 345	 & 1 054	 & 1 790	 & 4 679\\
Upper quartile pc 	 & 467	 & 656	 & 1 185	 & 1 771\\
median pc	 & 334	 & 523	 & 925	 & 1 268\\
Lower quartile pc 	 & 225	 & 359	 & 648	 & 947\\
min pc	 	 & 13	 & 73	 & 134	 & 196\\
\end{tabular}
\caption{Statistics for populations in RAVE identified as dwarfs, T-Tauri and subgiants (TT\&S), protostars and red giant branch (PS\&R), and red clump giants (RC).}
\end{flushleft}
\end{table}

I identified 21\,230 dwarfs and 13\,280 other classes by surface gravity and colour. The distance calibration can be reasonably be applied to a further 18 284 probable dwarfs for which no surface gravity measurement is available, but which have colour index $J-K \le 0.45$ mag, below which value very few stars are not main sequence dwarfs. Of those, RR Lyrae and $ \delta $-Scuti lie on the main sequence in the CMD, so that the distance calibration can be applied, and the number of Cepheids, luminous blue variables and Wolf-Rayet stars is not sufficient to impact properties of the velocity distribution. If the data base contained white dwarfs it would be possible to identify candidates because, having distance estimates too large by a factor of around 200, they would appear with unreasonable velocities on account of their proper motion, but white dwarfs do not appear to be represented in the RAVE data release because of their low luminosity. The resulting data base contains 52\,794 stars with statistics tabulated in table 4. 

In most cases the errors given in table 2 and table 3 are good estimates of the error in the calibration. These errors are dominated by real differences in stellar magnitude, given a population composed largely of field stars of all ages and compositions, and incorporate errors in 2MASS magnitudes (which are small). In addition, there will be an error due to the reddening correction. For 75\% of dwarfs the reduction in the distance estimate due to reddening is less than 7\%, and for 75\% of other classes it is less than 5\%. The maximum reduction in distance due to reddening is 34\% for dwarfs, 23\% for TT\&S, 19\% for PS\&R, but only 13\% for the red clump and 9.8\% for red dwarfs. Even in these worst cases, the error in these corrections will be dominated in quadrature by the errors in stellar magnitudes. An exception occurs when stellar class is incorrectly identified, which may cause a substantial error in distance.

\subsection{Solar velocity perpendicular to the Galactic plane}\label{sec:5.2}
I discarded any stars with MASK $ \le 0.7 $, below which value RAVE spectra are unreliable, and stars with any quality flags set, indicating issues which might affect the calculation of luminosity distance, such as multiple stars and peculiar objects. I excluded stars with luminosity distances greater than 1\,200 pc, for which proper motions are small and proper motion errors account for a substantial part of velocity estimates. These restrictions lead to a population of 24\,861 dwarfs, 4\,002 red clump giants, 1\,641 protostars and red giant branch, and 1\,828 T-Tauri and subgiants. 
\begin{figure}
	\centering
		\includegraphics[width=0.47\textwidth]{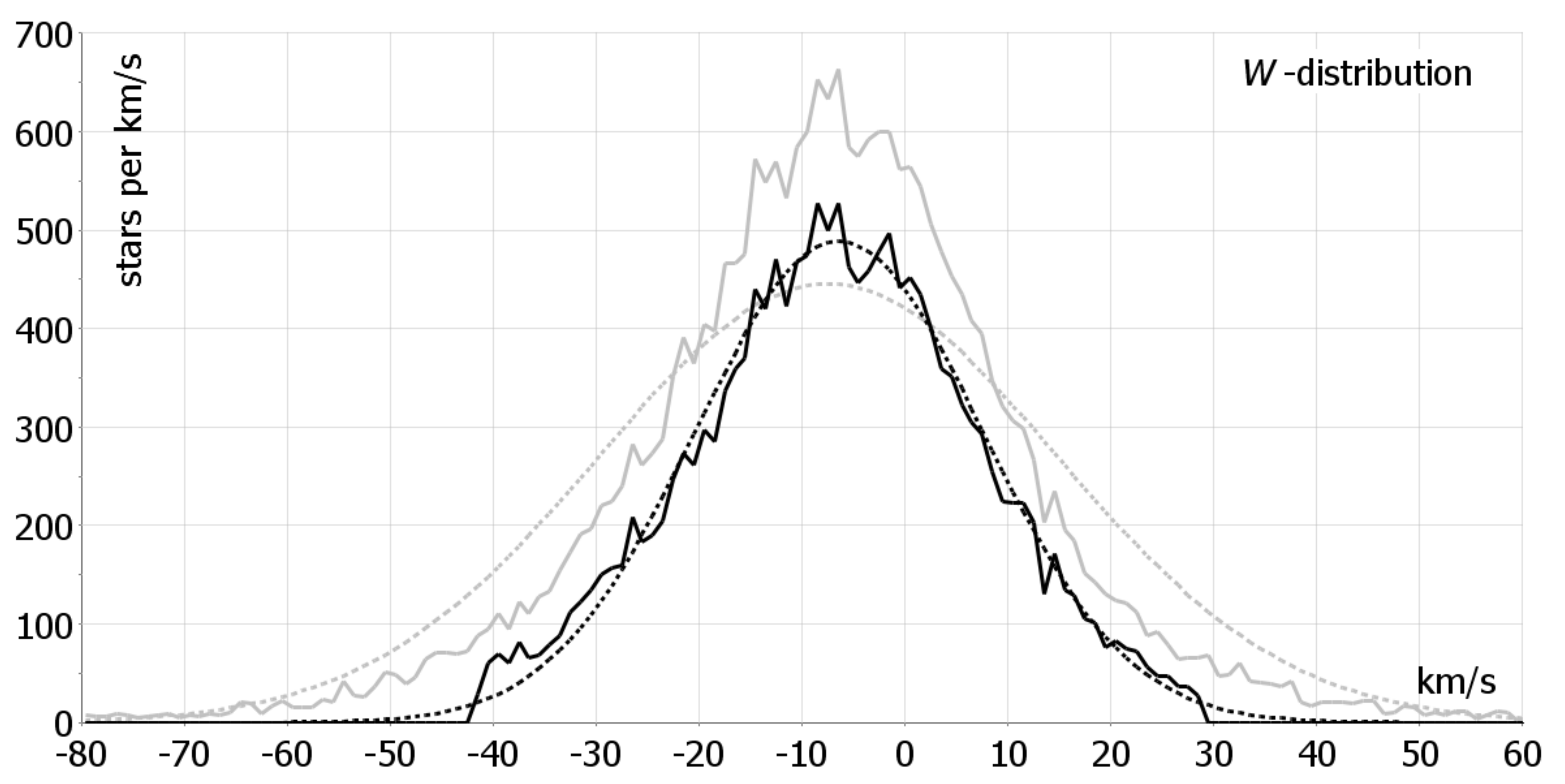}
		\caption{The velocity distribution perpendicular to the Galactic plane for the entire population of dwarfs (grey continuous) and after truncating to find a best fit with a Gaussian distribution (black continuous). Gaussian distributions with the same mean and standard deviation are shown in grey.}
	\label{Fig:8}
\end{figure}

Mean velocity for dwarfs is $ (\overline{U}, \overline{V}, \overline{W}) = (-12.8 \pm 0.2, -18.8 \pm 0.2, -7.46 \pm 0.14)$ km\,s$^{-1} $ with dispersion $ (\sigma_U, \sigma_V, \sigma_W) = (36.6, 29.7, 22.3) $ km\,s$^{-1}$, but these figures are influenced by fast moving stars, which are generally old stars and can be considered distinct from the main population of the thin disc. I restricted the population of dwarfs to stars with eccentricity in the $ UV $ plane less than $ 0.4 $, based on $ (U_0, V_0) = (14.1, 14.6) $ km\,s$^{-1}$. I restricted the population to stars with $ |W + W_{\mathrm{G}}| < w $, and varied $ W_{\mathrm{G}} $ and $ w $ so as to minimise the sum of least squares differences between the frequency distribution, in 1 km\,s$^{-1}$ bins, and a Gaussian distribution with mean $ W_{\mathrm{G}} $ and standard deviation equal to dispersion for the truncated population (figure 8). The best fit has $ W_{\mathrm{G}} = 6.52 \pm 0.05 $, $ w = 35 \pm 0.5 $ and dispersion $ \sigma_W = 13.9 $ km\,s$^{-1}$.

The best estimate the speed of Solar motion perpendicular to the Galactic plane is from the mean of the truncated distribution, $ W_0 = -\overline{W} = 6.89 \pm 0.11 $, in close agreement with Hipparcos, for which the best fit has $ W_{\mathrm{G}} = 6.9 \pm 0.1 $, $ w = 26 \pm 1 $, $ \sigma_W = 10.5$ and $ W_0 = -\overline{W} = 6.9 \pm 0.1 $ (XHIP). The higher dispersion in RAVE is partly due to the difference in cut-off $ w $ (with $ w = 35 $, dispersion for Hipparcos is $ 12.4 $ km\,s$^{-1}$), and partly because greater velocity errors arise from less accurate RAVE distances and proportionately greater proper motion errors.

Repeating the exercise for other classes in RAVE gave $ W_{\mathrm{G}} = 6.9 \pm 0.1$, $w = 56 \pm 1 $, $ \sigma_W = 23.7 $, and $ W_0 = -\overline{W} = 7.0 \pm 0.3 $ km\,s$^{-1}$ km\,s$^{-1}$, in agreement with the values from dwarfs and from Hipparcos. The much larger velocity dispersion results from the greater distances of these populations and proportionately greater errors in proper motion, and from the high proportion of thick disc stars.

\subsection{The thin and thick disc}\label{sec:5.3}
Based on the velocity dispersion, $ \sigma_W = 10.5 $ km\,s$^{-1}$, of the Gaussian approximation to the truncated distribution for Hipparcos stars (used here because Hipparcos kinematics are more accurate than RAVE kinematics), one can set a pragmatic $ 3\sigma $ bound, $ | W + 6.9 | \le 31 $ km\,s$^{-1}$, on velocities of thin disc stars at the Galactic plane. In a first approximation using planar motion, a star in the Galactic plane in a circular orbit with velocity 31 km\,s$^{-1}$ perpendicular to the plane will rise to a height of 1.1 kpc above the plane (using $R_0 = 7.4$ and velocity of circular motion 211 km\,s$^{-1}$) a factor of $ \sim$3 - 4 greater than a typical scale height, $ \sim $300 pc, given for the thin disc in the literature (e.g., Robin et al. 2003). This is an overestimate, but since the total gravitating mass of the inner Galaxy and halo greatly outweighs matter distributed in the disc at the solar radius, it may not be a substantial overestimate.

In a reasonable approximation within the thin disc, the vertical component of gravitational force may be taken as proportional to the height above the disc. Using an adopted height of 900 pc for the thin disc, and a height of 6pc for the Sun above the Galactic plane (consistent with observations of SgrA*), we may regard stars satisfying 
\begin{equation}\label{eq:5.3.1}
\frac{(Z+6)^2}{900} + \frac{(W+7)^2}{31} < 1
\end{equation}
as thin disc stars, and other stars as thick disc stars. This is not intended as a crisp boundary between distinct populations, but is rather a convenient division into bins with kinematically different properties (it is generally true that a boundary between bins is an arbitrary choice, not a physically extant boundary).
\begin{figure}
	\centering
		\includegraphics[width=0.47\textwidth]{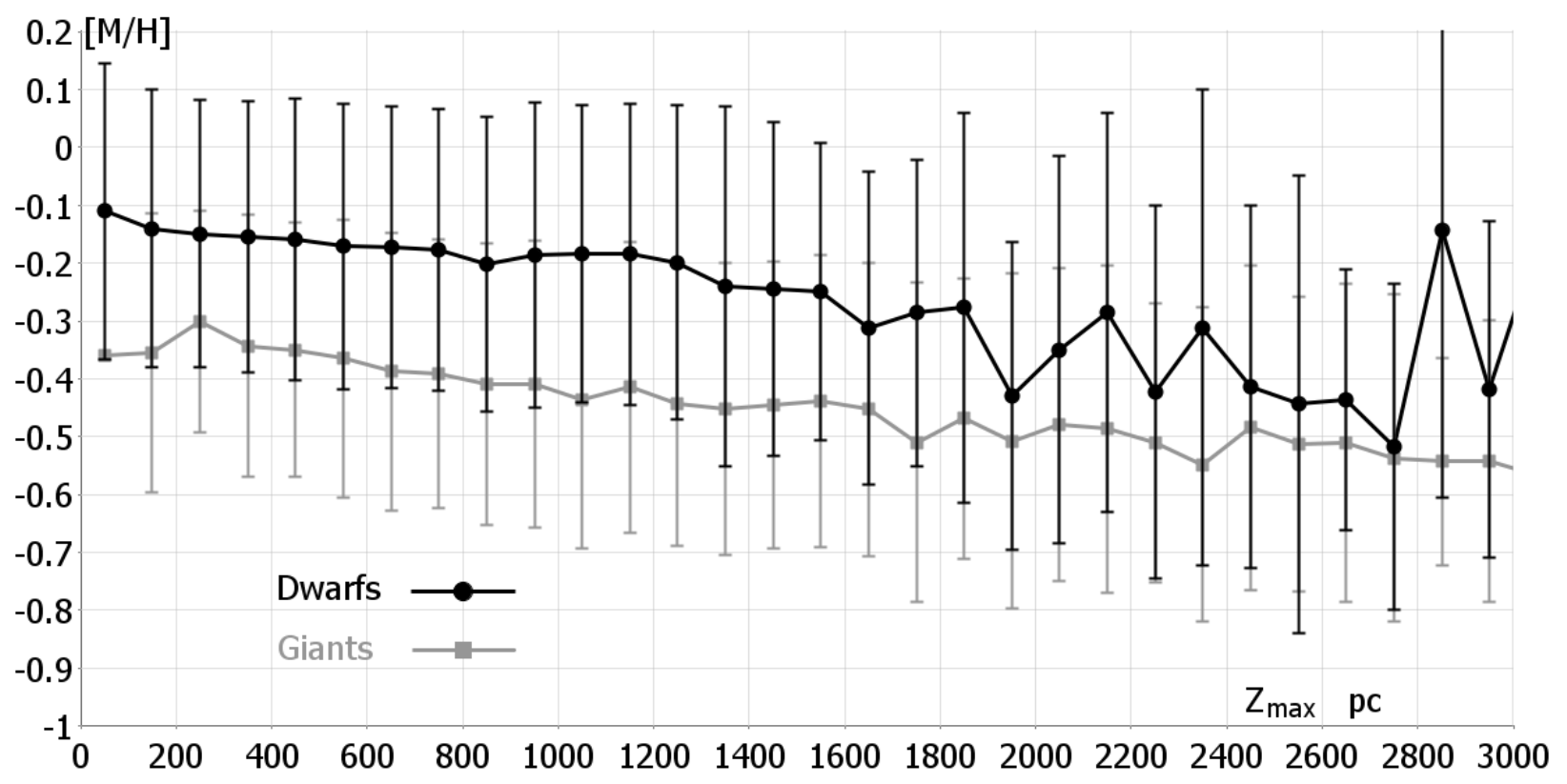}
		\includegraphics[width=0.47\textwidth]{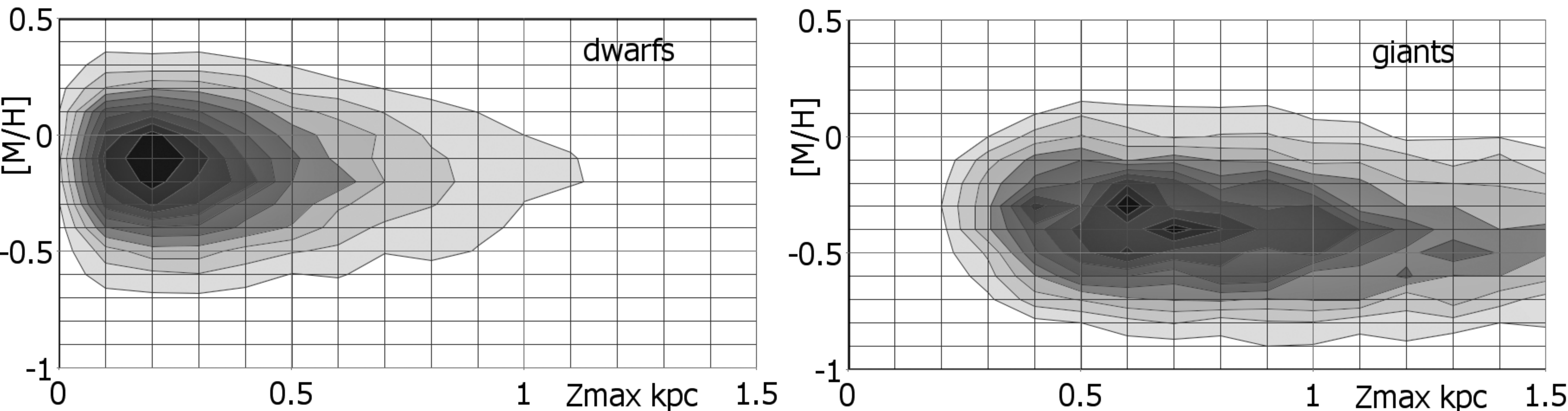}
		\caption{Top: mean metallicity of RAVE dwarfs and giants binned by (approximate) maximal orbital height above the Galactic plane. Error bars show the standard deviation for each bin. Bottom: the distribution of metallicities of RAVE dwarfs and giants plotted against maximal orbital height. The roughly constant gradient and unimodal form of the distribution supports the argument of Bovy et al. that there is no distinct division between thick and thin discs.}
	\label{Fig:9}
\end{figure}

Based on equation (\ref{eq:5.3.1}), the maximum height of a stellar orbit is given roughly by
\begin{equation}\label{eq:5.3.2}
Z_{\mathrm{max}}= \sqrt{(Z+6)^2 + (\frac{900}{31}(W+7))^2}
\end{equation}
I binned the population into 100pc bins by $ Z_{\mathrm{max}} $ and plotted mean metallicity for each bin (figure 9) for dwarfs and for giants. Error bars show standard deviation in each bin. A broad range of metallicities is seen at all heights. Systematic variations in metallicity with disc height are much less than random variations, so that, although metallicities are generally lower for thick disc stars, metallicity is not a good indicator of disc membership. The plots for both dwarfs and giants show approximately constant gradient with random fluctuation, and the distribution of metallicities against maximum disc height is unimodal. This supports the argument that there is no distinct division between thick and thin discs (Bovy et al. 2012). The absence of any sharp division opposes the idea that the thick disc was created by a sudden event such as galactic merger and supports the notion that it is a remnant of the much more chaotic early dynamics of the Galaxy and continuous evolution towards a cooler and less chaotic state, wherein younger stars were created with progressively nearer to planar orbits.

\subsection{Thin disc kinematics}\label{sec:5.4}
Smoothing of velocity distributions (figures 10-13 \& 15-16) is obtained by representing each data point as a 2D Gaussian centred at that point with standard deviation in each axis given by the smoothing parameter. The smoothing parameter gives a direct measure of the `blurring' of the distribution used to show density and obscure random artefacts. For best accuracy, the smoothing parameter should be as low as possible consistent with a clear image so as to remove, in so far as is possible, random variations while leaving structure intact. In general, the greater the density of stars in the interesting central part of the distribution, the lower the smoothing parameter which can be used.
\begin{figure}
	\centering
		\includegraphics[width=0.47\textwidth]{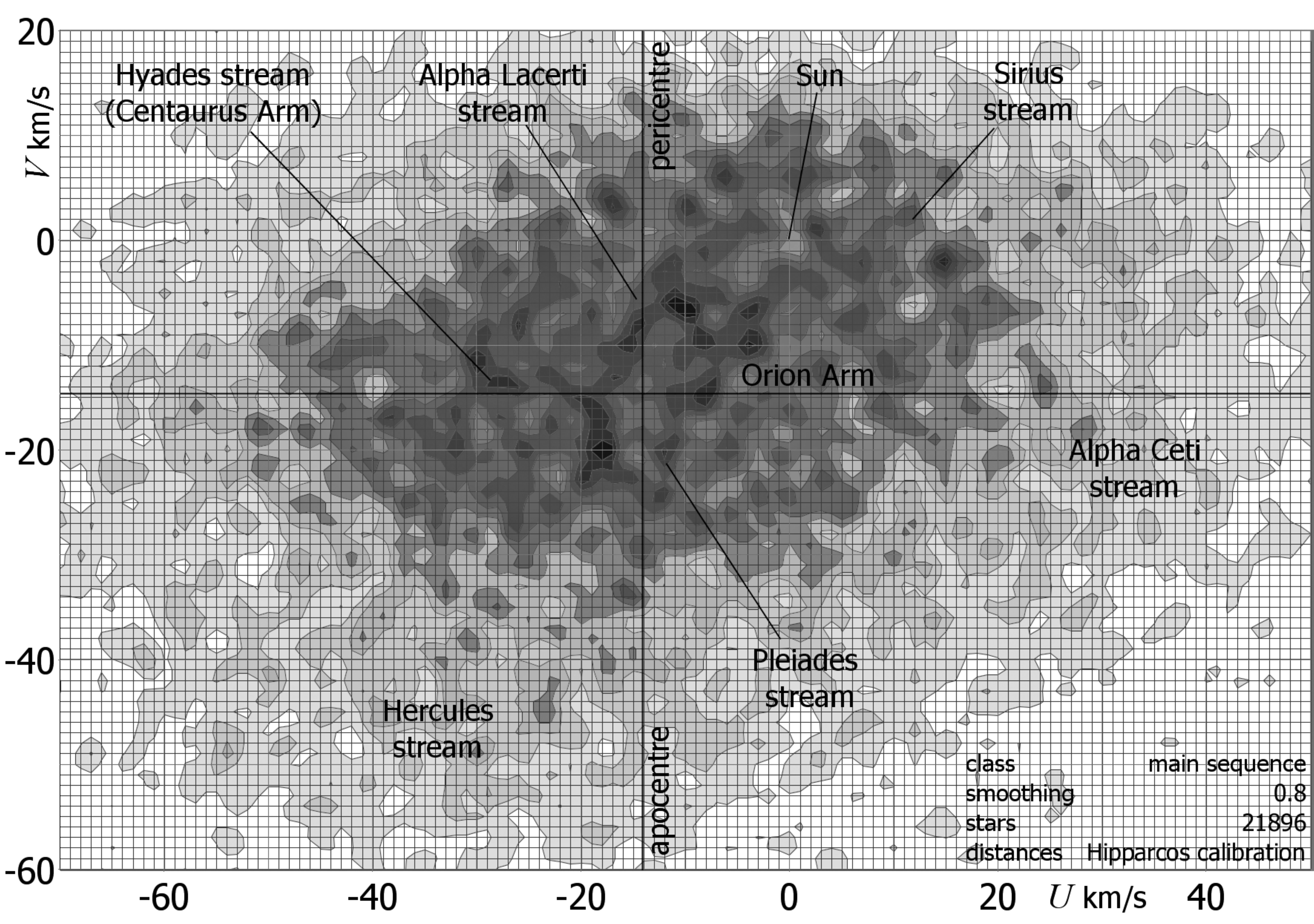}
		\caption{The distribution of $ U $- and $ V $-velocities for RAVE thin disc dwarfs with distances calibrated to Hipparcos. Positions of the Hyades, Sirius, Hercules, Alpha Lacertae and Alpha Ceti streams are shown. The Pleiades stream is notably reduced, or displaced, because of the greater distance of these stars and the uneven distribution of star formation.}
	\label{Fig:10}
\end{figure}

The velocity plot for main sequence stars (figure 10) shows a number of features previously seen in the $ UV $-distribution for Hipparcos stars (Francis \& Anderson 2012, `FA12', figure 8). The Hyades, Sirius, Hercules, Alpha Lacertae and Alpha Ceti streams are all observed, as is the well in the velocity distribution close to the LSR. These streams are also observed in RAVE by Antoja et al. (2012). Demarcation between streams is less clear than in the Hipparcos data because there are greater variations in the velocity distribution over the greater distances covered by RAVE data, and because measurement is less accurate. The demarcation will be seen with greater clarity in section 5.8 from the $ e $-$\phi$ plot (figure 21). Note that streams are broad motions; individual peaks are likely to be caused by random fluctuation. 

The Pleiades stream is notably reduced or displaced, illustrating that the intensity and peculiar motions of star formation vary with position. The RAVE distribution contains a greater proportion than Hipparcos of stars in the Orion arm with velocities to the right of the LSR on the diagram. This is as predicted by the model of spiral structure described in Francis and Anderson (2009b), as stars with these motions are expected further from the Solar position toward the inside of the arm. There is a hole in the distribution of the Hyades stream, at close to the position of the peak value in the Hipparcos distribution. This is explained because stars from the Centaurus arm, cross the Orion arm at a particular point in their orbit; in the immediate solar neighbourhood stars from the Centaurus arm have velocities centred at a particular value. Stars at a greater distance from the Sun have velocities displaced from that value.
\begin{figure}
	\centering
		\includegraphics[width=0.47\textwidth]{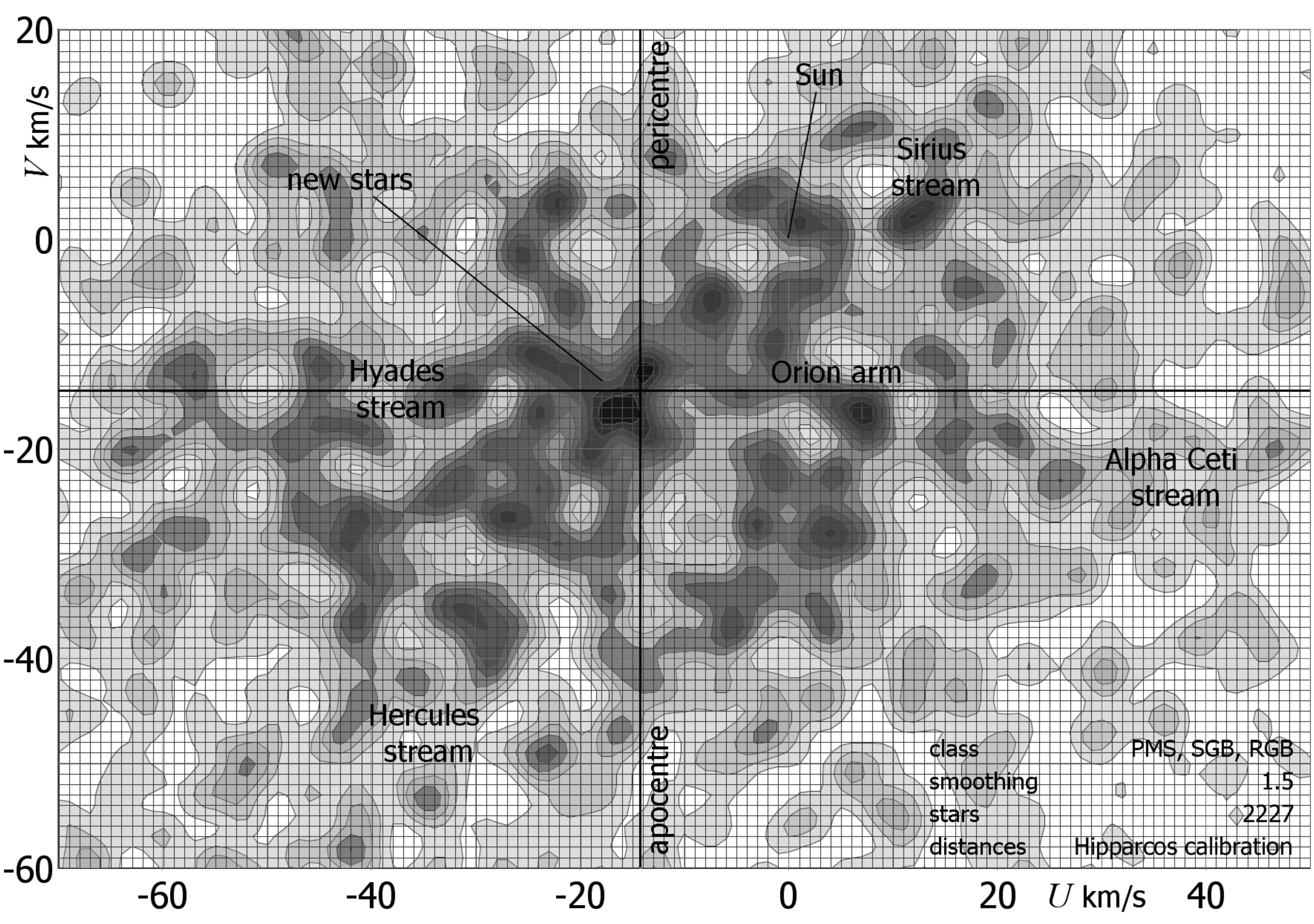}
		\caption{The velocity distribution for pre-main sequence stars, subgiants and red giant branch in the thin disc. Axes are shown with an origin at the LSR, $ (U_0, V_0) = (14.1, 14.6) $ km s$ ^{-1} $ (XHIP).}
	\label{Fig:11}
\end{figure}

The velocity plots for T-Tauri and subgiants (TT\&S) and for protostars and red giant branch (PS\&R), show that these populations contain substantial numbers of new stars, with motions close to the LSR. These populations are not large and I have combined them into a single population of pre-main sequence stars (PMS), subgiants and red giant branch to reduce the effect of random errors and obtain a better view of kinematics (figure 11). There is a significant scatter in the velocity plot, largely due to subgiants and the RGB. It will be observed that this group is most prone to misidentification of stellar class, which may generate substantial distance errors for particular stars. Nonetheless, the Hyades, Sirius, Alpha Ceti, Hercules streams are visible, and there is a broad representation of stars in the Orion arm at greater distances from those seen in Hipparcos.
\begin{figure}
	\centering
		\includegraphics[width=0.47\textwidth]{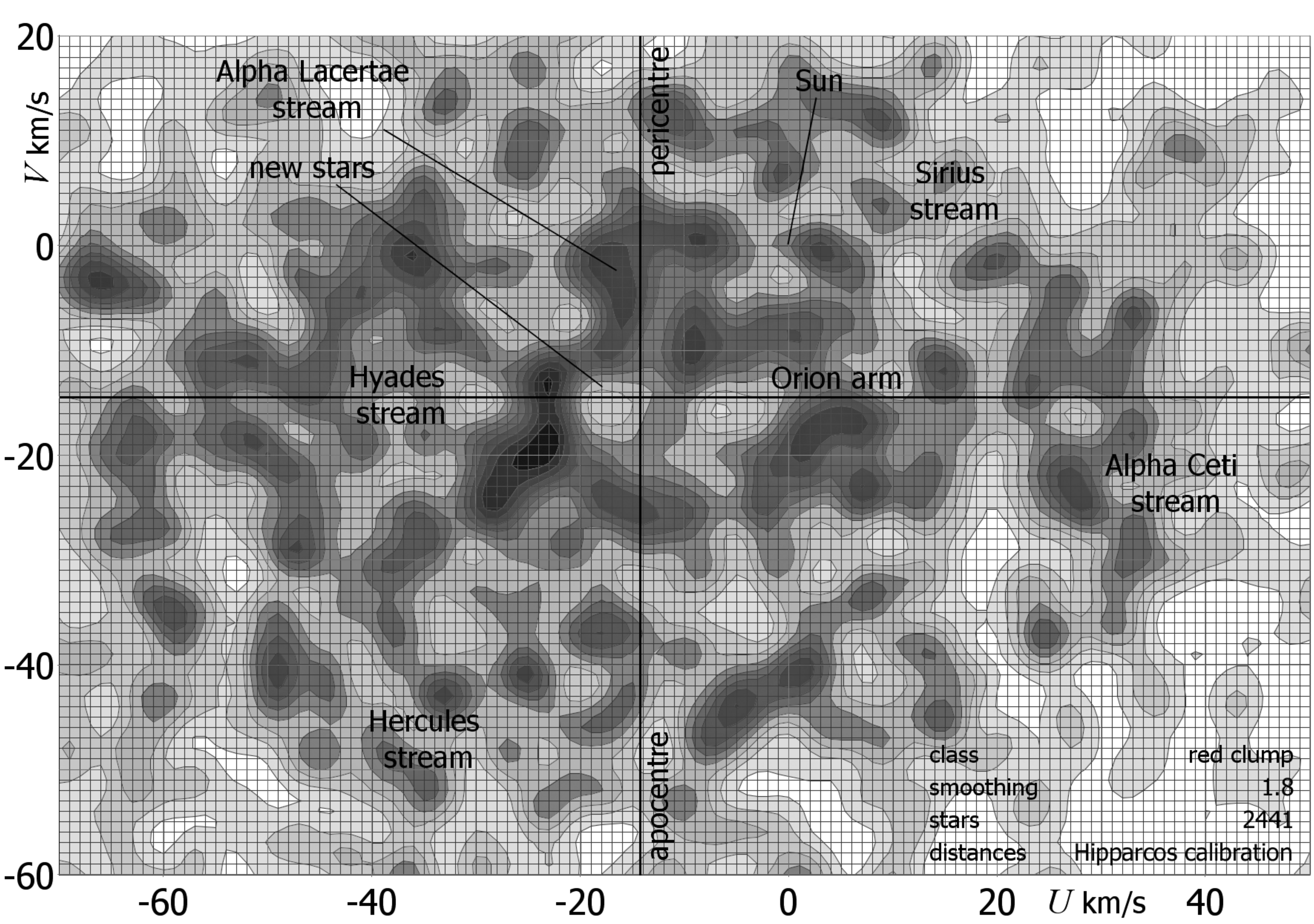}
		\caption{The distribution of $ U $- and V-velocities for thin disc RAVE giants with distances up to 1\,200 pc. from the Hipparcos calibration. }
	\label{Fig:12}
\end{figure}

The distribution of thin disc giants (figure 12) also shows the major streams. The Hyades, Sirius, Alpha Ceti, Hercules and Alpha Lacertae streams are visible. The velocities of more distant stars in the Orion arm are well represented. Since giants are mature stars, the absence of velocities typical of new stars illustrates again that stellar orbits evolve away from circular motion. The hole in the distribution of the Hyades stream, seen previously for dwarfs, is pronounced, due to the greater distances of giants. 
\begin{figure}
	\centering
		\includegraphics[width=0.47\textwidth]{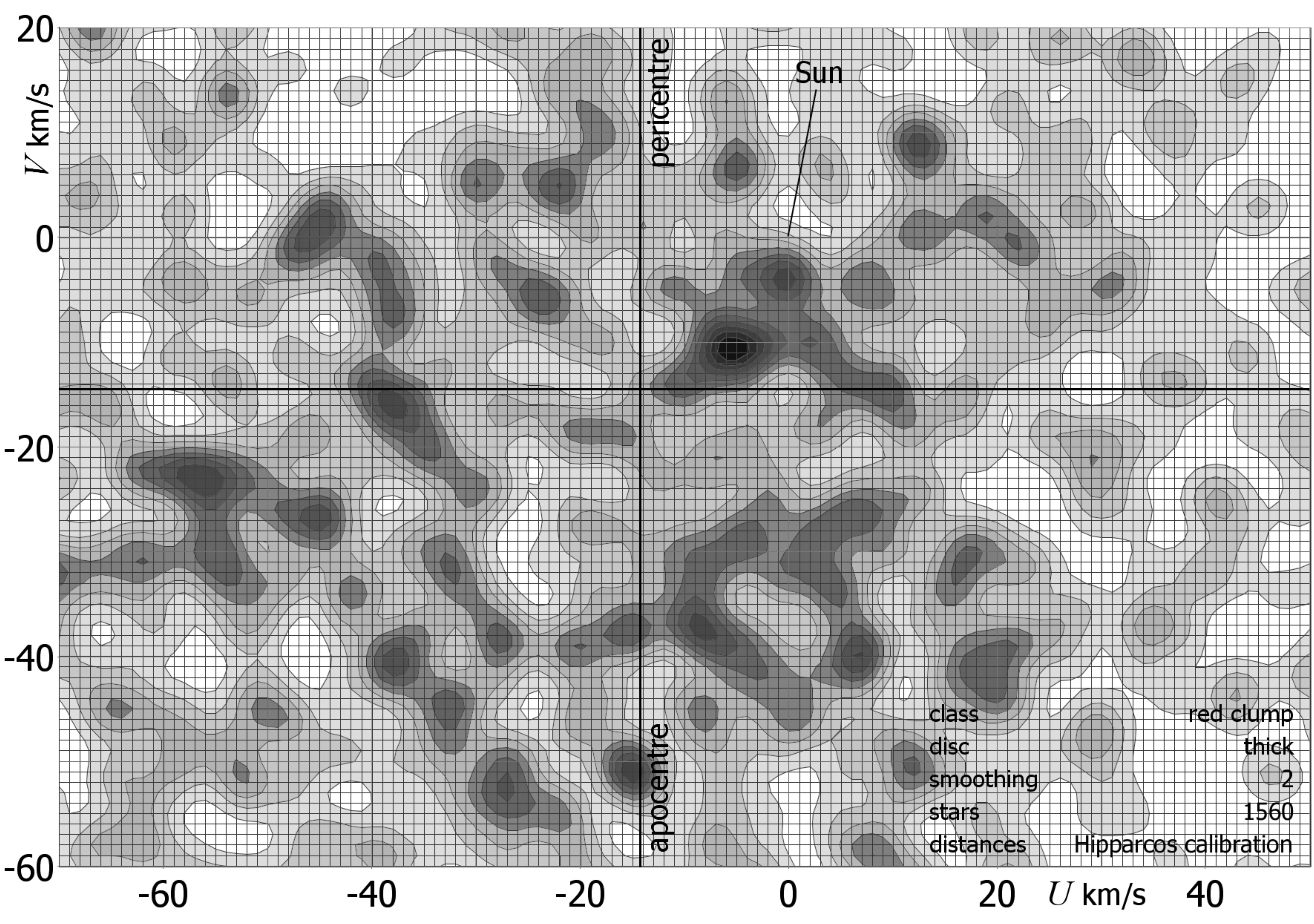}
		\caption{The distribution of $ U $- and V-velocities for thick disc RAVE giants with distances up to 1\,200 pc. from the Hipparcos calibration. }
	\label{Fig:13}
\end{figure}

\subsection{Thick disc kinematics}\label{sec:5.5}
The velocity distribution of thick disc RAVE giants, with distances within 1200 pc from the Hipparcos calibration (figure 13) shows a greater velocity dispersion and absence of definite structure in the thick disc. A substantial peak is seen at the position of the Orion arm, and consists of stars which might be more correctly classified in the thin disc. Other peaks which I tested are simply random fluctuations, not moving groups, as they do not show localisation in space or particular values of $W$ km\,s$^{-1}$.
\onecolumn
\begin{figure}
	\centering
		\includegraphics[width=1\textwidth]{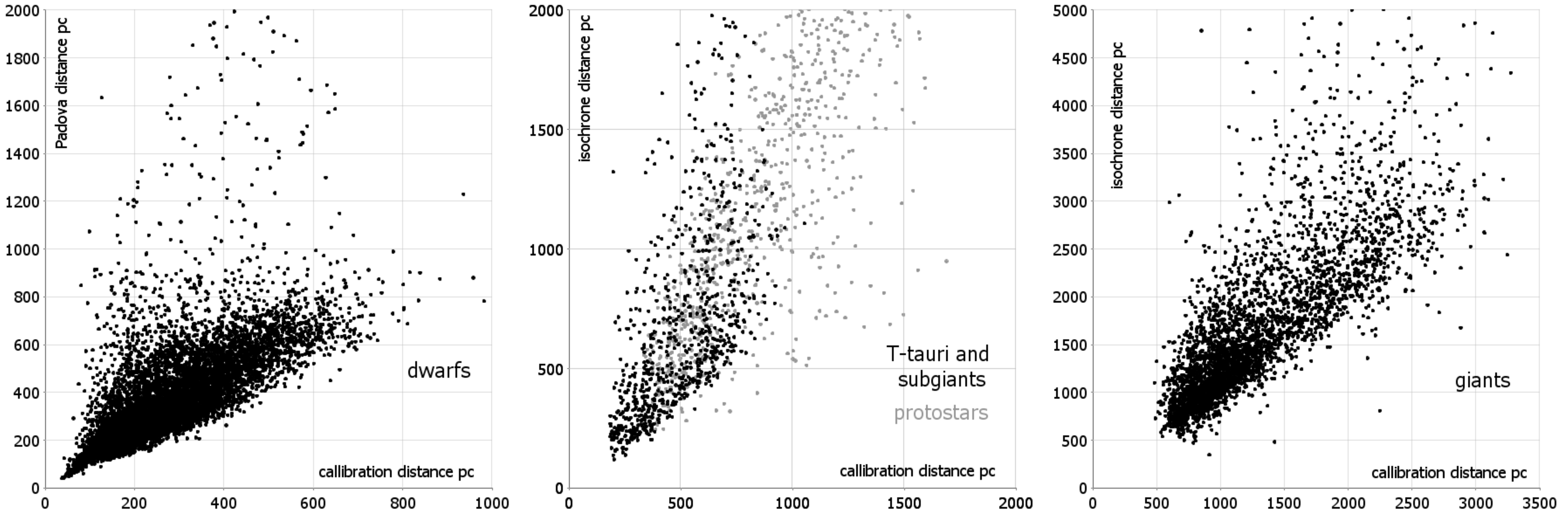}
		\caption{Comparison of isochrone distances with distances as calibrated in this paper. The scatter of points with greater isochrone distances in each plot can be attributed to the misidentification of stellar classes.}
	\label{Fig:14}
\end{figure}
\begin{table}\label{(table5)}
\begin{flushleft}
\begin{tabular}{lrrrrrrrr} 
\textbf{class} &\multicolumn{3}{c}{ \textbf{ thin disc dwarfs}} & \multicolumn{3}{c}{\textbf{thin disc RC}} & \multicolumn{2}{c}{\textbf{thick disc RC} } \\ 
\textbf{distances}	 & \textbf{Hip}	 & \textbf{calib}	 & \textbf{Padova}	 & \textbf{Hip}	 & \textbf{calib}	 & \textbf{Padova}	 & \textbf{calib}	 & \textbf{Padova}\\
stars	 & $ 13\,468	 $ & $ 22\,107	 $ & $ 7\,215	 $ & $ 5\,540	 $ & $ 2\,439	 $ & $ 841	 $ & $ 1\,560	 $ & $ 533$\\
$ \overline{U} $ km\,s$^{-1}$	 & $ -9.9	 $ & $ -12.1	 $ & $ -14.2	 $ & $ -8.3	 $ & $ -21.3	 $ & $ -24.2	 $ & $ -25.8	 $ & $ -33.0$\\
$ \sigma_{U} $ km\,s$^{-1}$	 & $ 30.0	 $ & $ 32.6	 $ & $ 40.5	 $ & $ 32.3	 $ & $ 47.1	 $ & $ 51.0	 $ & $ 61.8	 $ & $ 65.9$\\
$ \sigma_{\overline{U}} $ km\,s$^{-1}$	& $ 0.26	 $ & $ 0.22	 $ & $ 0.48	 $ & $ 0.43	 $ & $ 0.95	 $ & $ 1.8	 $ & $ 1.6	 $ & $ 2.9$\\
$ \overline{V} $ km\,s$^{-1}$	 & $ -16.5	 $ & $ -16.8	 $ & $ -20.3	 $ & $ -17.8	 $ & $ -25.6	 $ & $ -28.2	 $ & $ -42.6	 $ & $ -42.9$\\
$ \sigma_{V} $ km\,s$^{-1}$ & $ 20.4	 $ & $ 22.8	 $ & $ 28.3	 $ & $ 22.2	 $ & $ 34.0	 $ & $ 33.9	 $ & $ 49.0	 $ & $ 48.3$\\
$ \sigma_{\overline{V}} $ km\,s$^{-1}$	 & $ 0.18	 $ & $ 0.15	 $ & $ 0.30	 $ & $ 0.30	 $ & $ 0.69	 $ & $ 1.2	 $ & $ 1.2	 $ & $ 2.1$\\
$ \overline{W} $ km\,s$^{-1}$	 & $ -7.1	 $ & $ -6.8	 $ & $ -7.1	 $ & $ -7.0	 $ & $ -7.1	 $ & $ -6.3	 $ & $ -9.2	 $ & $ -10.8$\\
$ \sigma_{W} $ km\,s$^{-1}$	 & $ 11.2	 $ & $ 13.2	 $ & $ 13.6	 $ & $ 12.7	 $ & $ 13.6	 $ & $ 13.6	 $ & $ 48.3	 $ & $ 49.3$\\
$ \sigma_{\overline{U}} $ km\,s$^{-1}$	 & $ 0.10	 $ & $ 0.09	 $ & $ 0.16	 $ & $ 0.17	 $ & $ 0.27	 $ & $ 0.45	 $ & $ 1.2	 $ & $ 2.1$\\
\end{tabular}
\caption{Statistics for populations in RAVE identified as dwarfs, T-Tauri and subgiants (TT\&S), protostars and red giant branch (PS\&R), and red clump giants (RC).}
\end{flushleft}
\end{table}
\twocolumn

\subsection{Comparison of statistics}\label{sec:5.6}
I plotted distances calculated by Zwitter et al. (2010) using Padova isochrones against distances from the Hipparcos calibration for stars in both data bases (figure 14). Other isochrones used by Zwitter et al. (2010) appear to give similar distances, but Padova isochrones have the advantage that they apply to giants as well as to dwarfs. For each class, the calibration has systematically smaller distances. There is also a scatter of stars with substantially greater isochrone distances in each plot. This can be attributed to differences between the identification of stellar class using surface gravity and colour, and the method used by Zwitter et al. (2010). Dwarfs wrongly identified as giants, and red clump giants wrongly identified as asymptotic branch giants will have greatly overestimated distances, and consequently greater velocities. The average distance for stars in both data bases is 28\% less than given by Zwitter et al. (2010) for 8\,748 dwarfs and 26\% less for 5\,355 members of other classes, but if the scatter of points showing misidentification of stellar class is ignored, the difference seen in the plots for both dwarfs and giants is nearer to 10\%.

I identified populations of thin disc dwarfs and giants among Hipparcos stars, and populations of thin disc dwarfs, thin disc red clump giants and thick disc red clump giants for stars with distances by Padova isochrones. I tabulated velocity statistics for each population in table 5. We clearly expect the stellar velocity distribution to vary on Galactic scales, and much less variation within a few hundred pc, so that there should be a close correspondence between kinematics of dwarfs in RAVE and Hipparcos. The exception is the Pleiades stream, which consists largely of young stars whose distribution in both space and velocity depends on the uneven distribution of star formation. A better match between RAVE and Hipparcos statistics for dwarfs is found using calibrated distances than using isochrone distances, and dispersions are less using the distance calibration than using isochrone distances. This is most naturally attributed to the better accuracy of the distance calibration.
\begin{figure}
	\centering
		\includegraphics[width=0.47\textwidth]{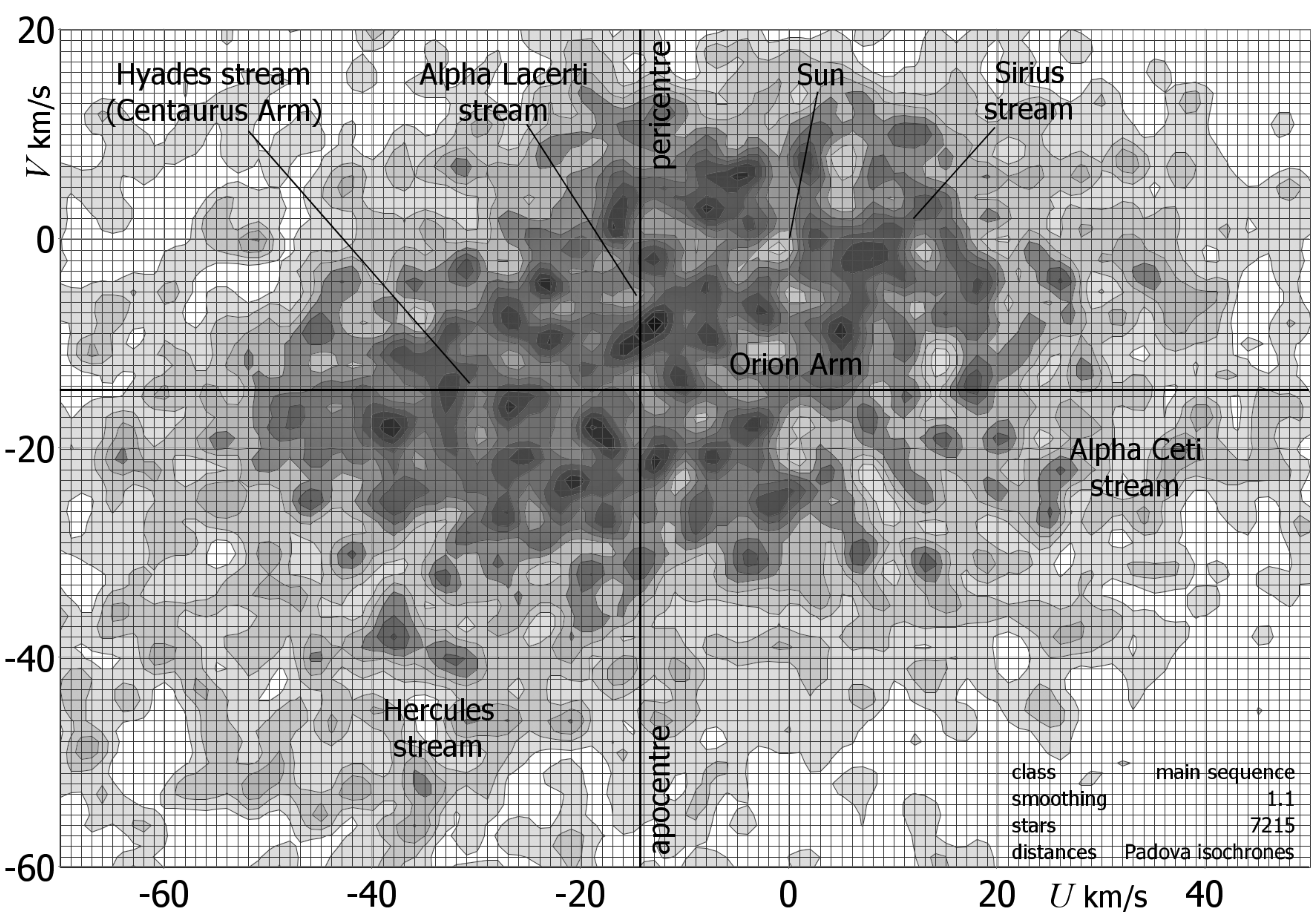}
		\caption{The distribution of $ U $- and $ V $-velocities for RAVE thin disc dwarfs with distances from Padova isochrones (c.f. figure 10). }
	\label{Fig:15}
\end{figure}
\begin{figure}
	\centering
		\includegraphics[width=0.47\textwidth]{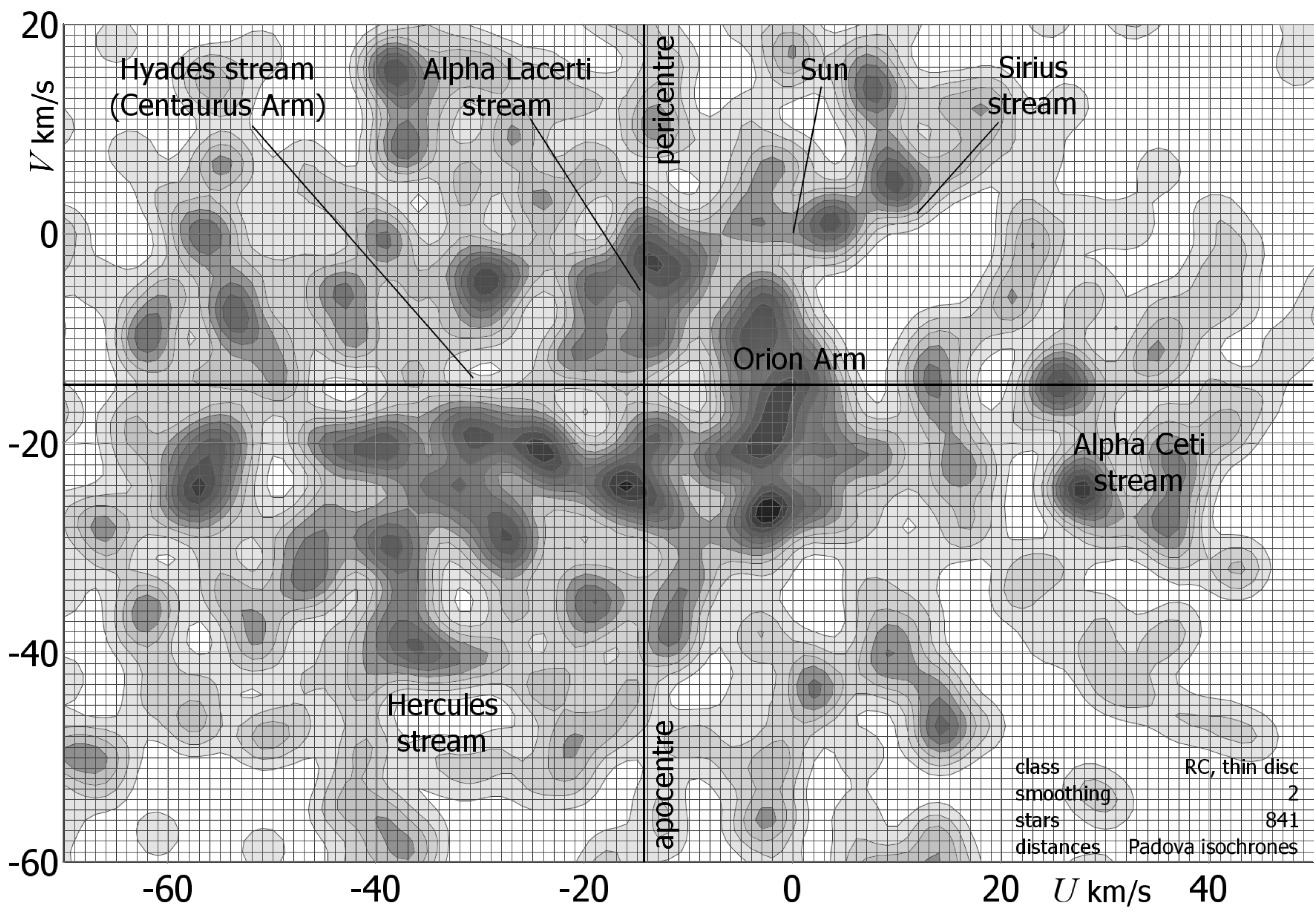}
		\caption{The distribution of $ U $- and $ V $-velocities for thin disc RAVE giants with distances up to 1\,200 pc from Padova isochrones (c.f. figure 12). }
	\label{Fig:16}
\end{figure}

Examination of the velocity plots is perhaps the best means to assess the accuracy of the data bases. The velocity distribution of RAVE dwarfs with distances calculated by Padova isochrones (figure 15) shows features similar to those found in the distribution with distances calibrated to Hipparcos (figure 10), including the major streams and the well at the position of the LSR, but it is less compact, showing greater dispersion, and the plot appears to contain more random variations in density, which may be put down to less accurate measurement of distance.
\begin{figure}
	\centering
		\includegraphics[width=0.47\textwidth]{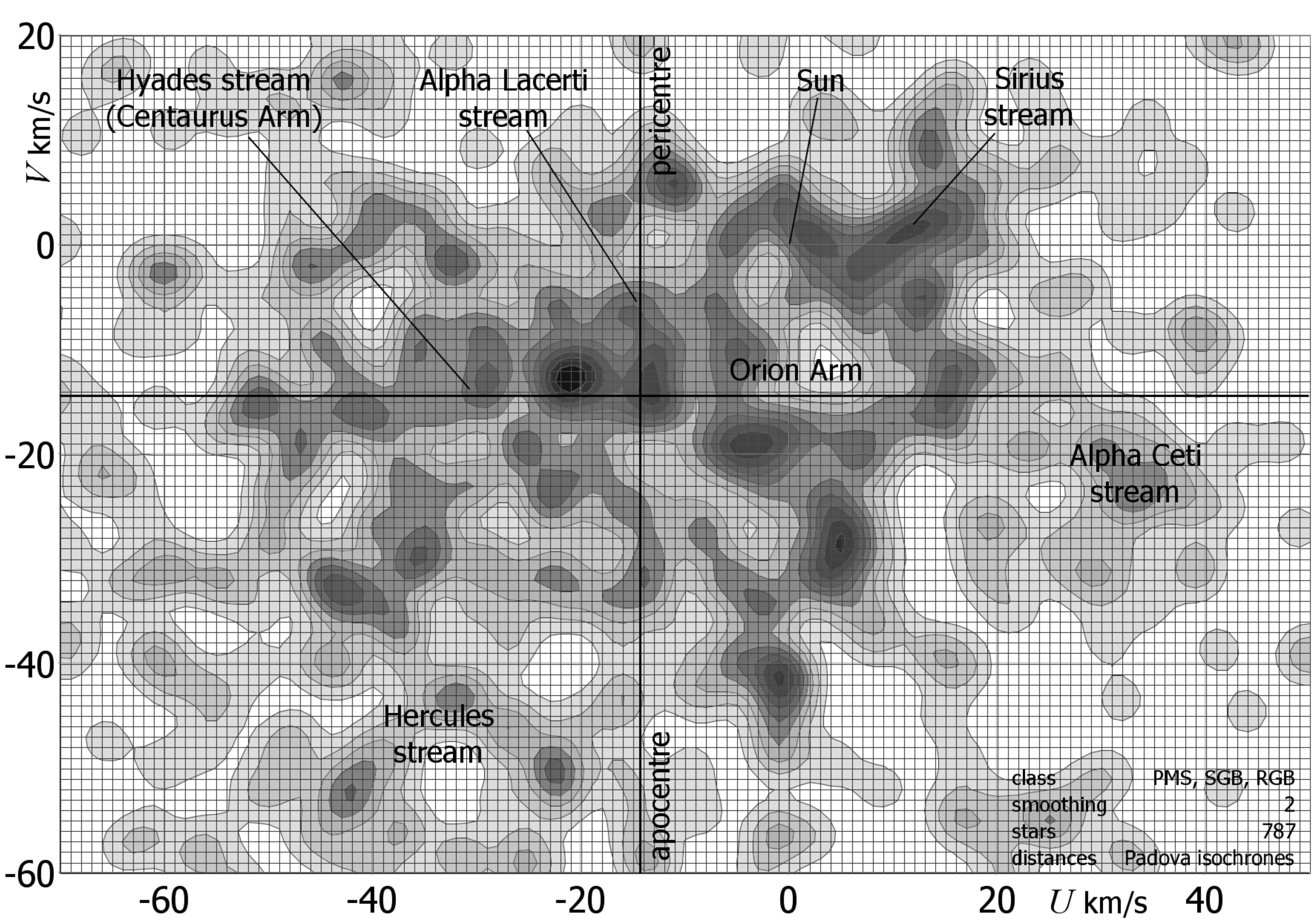}
		\caption{The velocity distribution for pre-main sequence stars, subgiants and the red giant branch in the thin disc, with distances from Padova isochrones (c.f. figure 11). }
	\label{Fig:17}
\end{figure}
\begin{figure}
	\centering
		\includegraphics[width=0.47\textwidth]{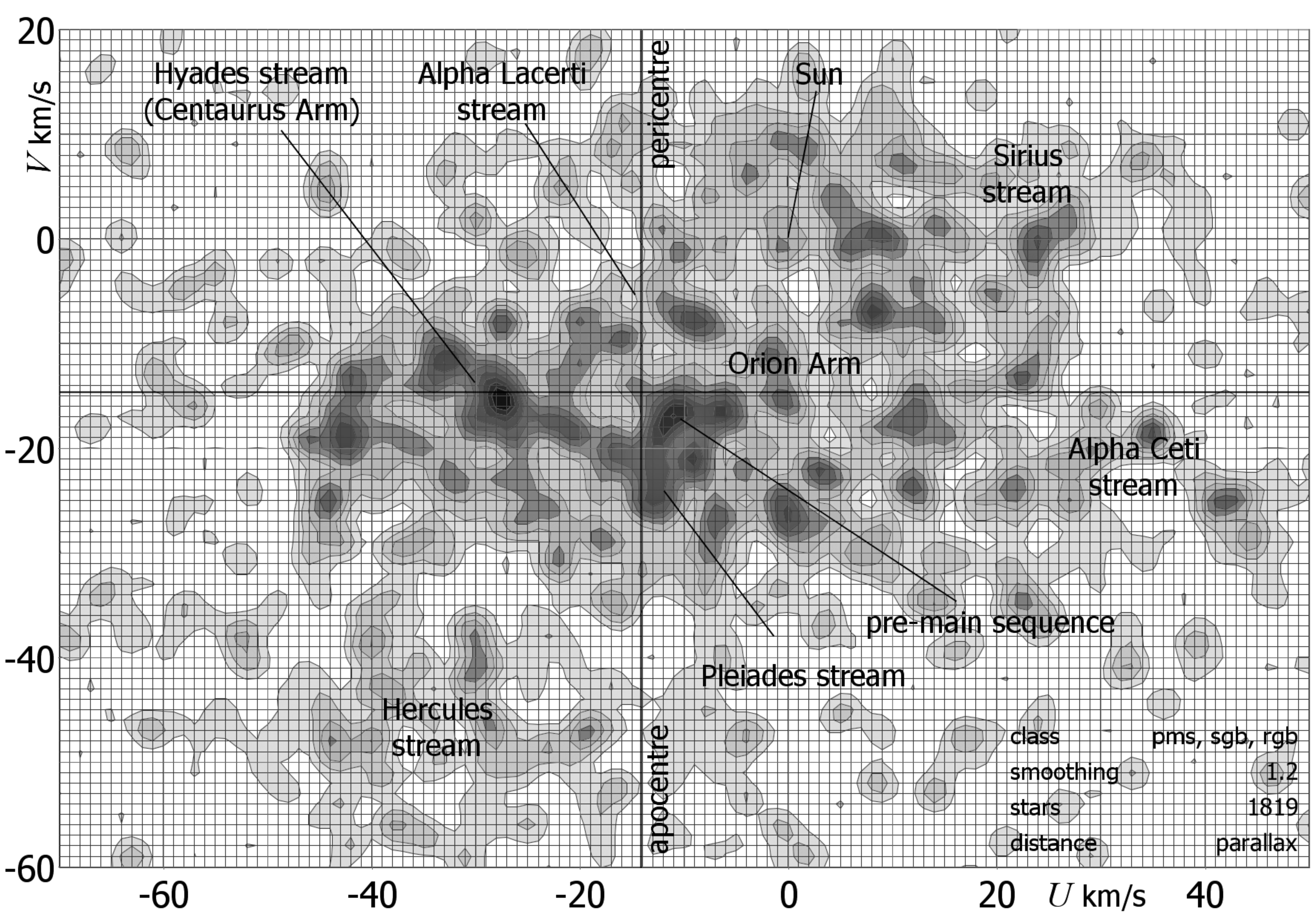}
		\caption{The velocity distribution for Hipparcos thin disc pre-main sequence stars, subgiants and the red giant branch with parallax distances (c.f. figure 11 and figure 17). Clusters are removed, but associations are not. }
	\label{Fig:18}
\end{figure}

The velocity distribution for red clump stars in the thin disc with isochrone distances (figure 16) also shows the major streams, with particularly strong representation of the Orion arm, as well as the holes in the Hyades and at the LSR. For red clump giants, the RAVE populations are substantially more distant, and table 5 shows a much greater disparity between RAVE and Hipparcos statistics within the thin disc. For both thin and thick disc giants velocities and velocity dispersions are typically a little higher with isochrone distances than with distances from the calibration. The thick disc naturally lags the Solar velocity more than the thin disc, but velocities away from the Galactic centre appear unrealistically high with both sets of distances, for both thin and thick disc giants. If proper motions are not systematically overstated for distant stars, this suggests that distances by both methods are high. If so, Padova isochrone distances are less accurate than the Hipparcos calibration.

As is the case for the Hipparcos calibration, the population of pre-main sequence stars, subgiants and red giant branch with distances by Padova isochrones shows significant motions close to the LSR (figure 17). Again there is some scatter, which may be attributed to subgiants and to the red giant branch. The major streams are visible. Figure 11 and figure 17 may be compared to the velocity distribution for pre-main sequence stars, subgiants and the red giant branch using Hipparcos parallax distances (figure 18), which also shows substantial motion close to the LSR.

\subsection{The local standard of rest}\label{sec:5.7}
Because of the strength of streaming motions, it is not possible to use mean velocities of thin disc stars to ascertain the local standard of rest. It is possible in principle to use thick disc motions, but proper motions of giants are small and errors in proper motion proportionately large. The magnitudes of $ \overline{U} $ for both the Hipparcos calibration and Padova isochrones (table 5) seem unreasonably high, and it is natural to attribute this to errors arising from large distances and small proper motions. 
\begin{figure}
	\centering
		\includegraphics[width=0.47\textwidth]{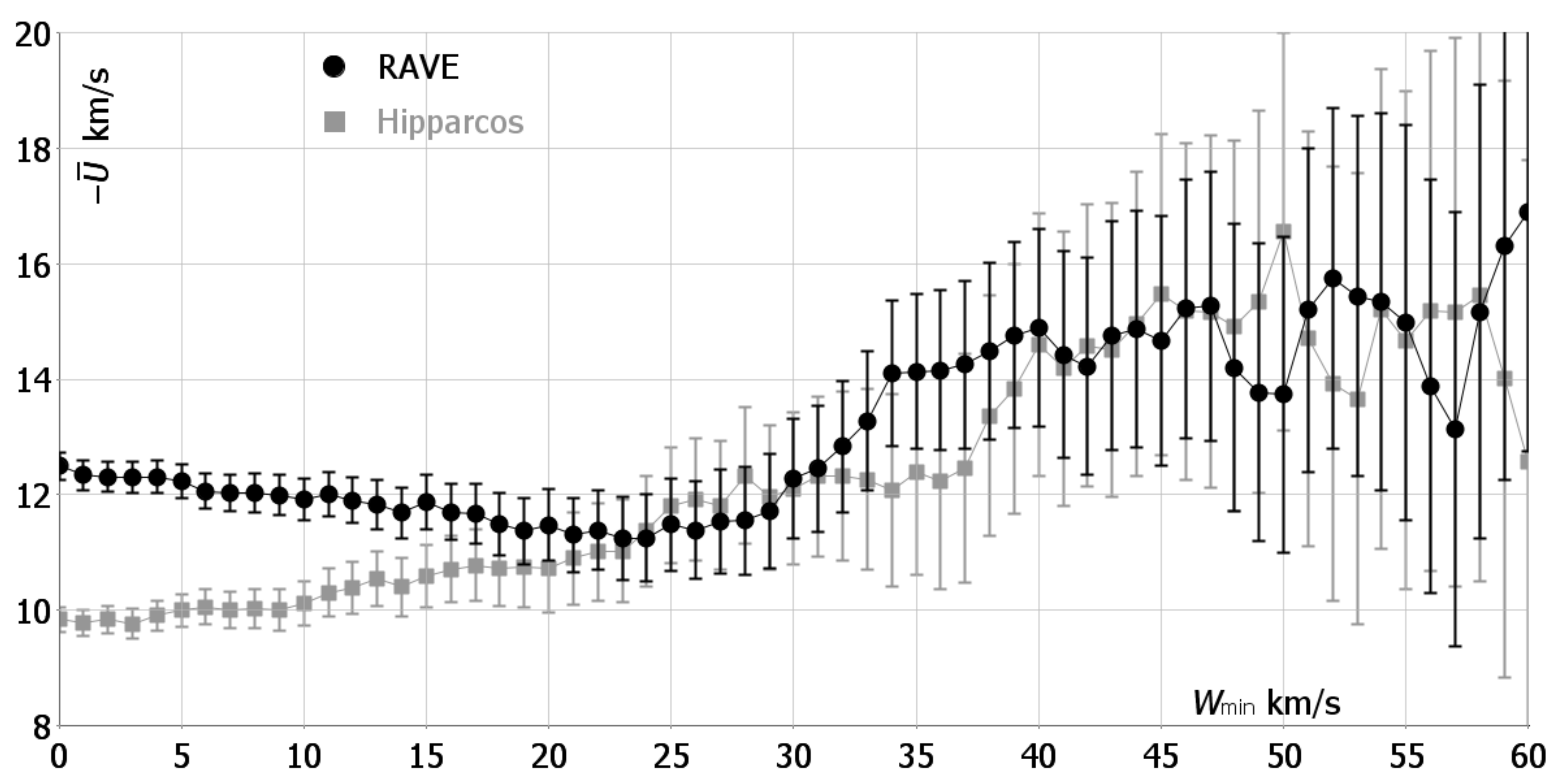}
		\caption{Mean $ U $ velocities, for stars with $ |W-7|>W_{\mathrm{min}} $ km\,s$^{-1}$. RAVE is shown with black circles. Hipparcos is shown with grey squares.}
	\label{Fig:19}
\end{figure}

To establish a population of thick disc stars unaffected by streaming motions, I applied a cut, $ W_{\mathrm{min}} $ km\,s$^{-1}$, on velocity perpendicular to the Galactic plane, removing stars with $ | W + 6.9 | < W_{\mathrm{min}} $ km\,s$^{-1}$. I excluded stars with velocities greater than 350 km\,s$^{-1}$ relative to the Galactic centre, which may be erroneous and which would have a disproportionate effect on the mean. I plotted $ |\overline{U}| $ against $ W_{\mathrm{min}} $ for main sequence RAVE stars with Hipparcos-calibrated distances (figure 19, black circles), and for Hipparcos stars (grey squares). There is a difference in mean radial velocity for low values of the cut-off, because the populations contain slightly different proportions of the main kinematic groups (section 5.8), and, because the RAVE population is more distant, RAVE stars in the Orion arm have greater velocity relative to the Sun. As the cut-off increases, the mean values of $U$ come together, levelling off with random fluctuations above $ W_{\mathrm{min}} = 40$ km\,s$^{-1}$, at which point orbits are sufficiently inclined to the Galactic plane that the influence of streaming motions is not significant. Using $ W_{\mathrm{min}} = 40$ km\,s$^{-1}$, gives an estimate of the LSR, with excellent agreement between the two data bases: from Hipparcos $ U_{0} = 14.6 \pm 2.3 $ km\,s$^{-1}$, and from RAVE $ U_{0} = 14.9 \pm 1.7 $ km\,s$^{-1}$.
\begin{figure}
	\centering
		\includegraphics[width=0.47\textwidth]{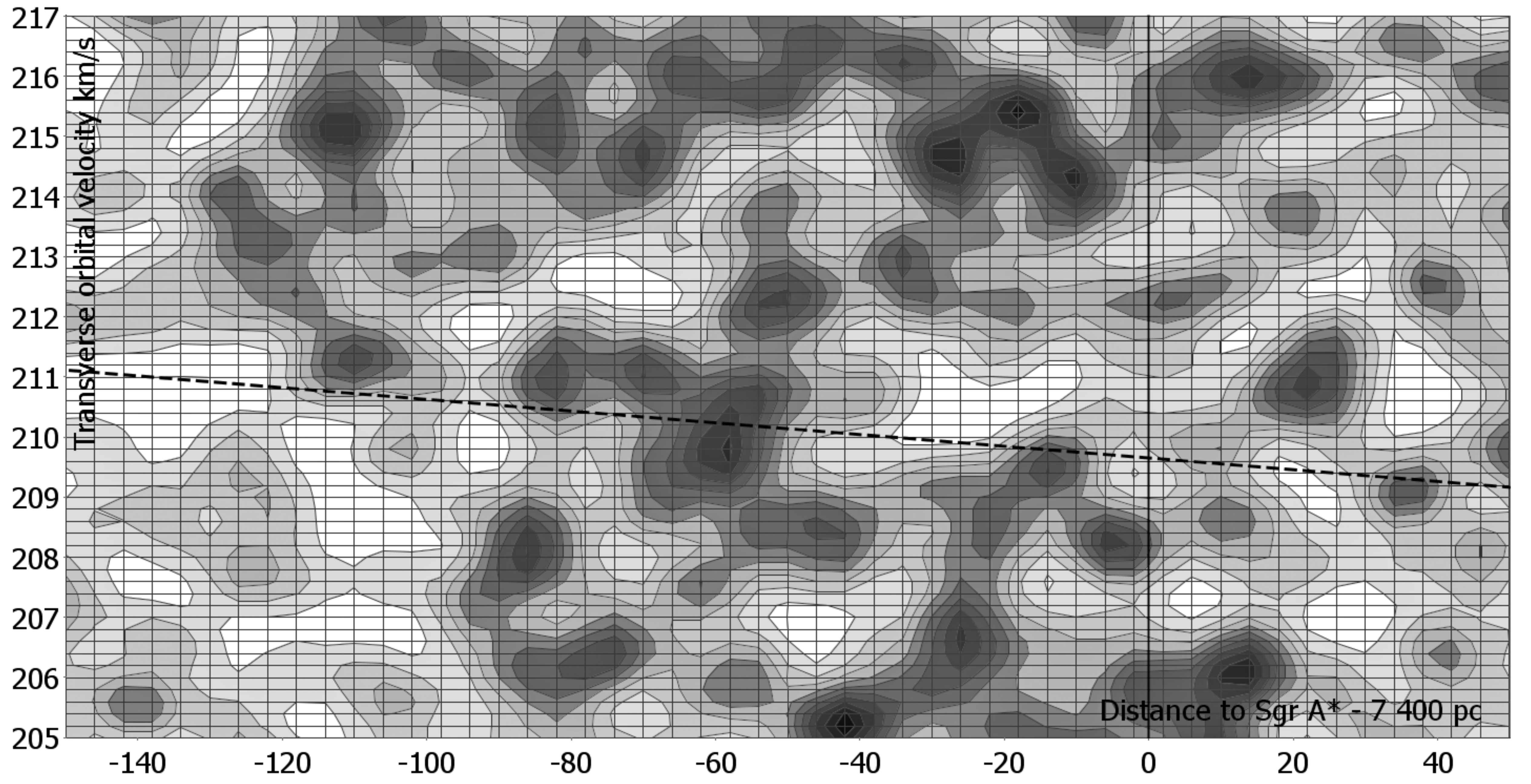}
		\caption{The transverse velocity distribution against distance to the Galactic centre, for stars with $ | U + 14 | < 8 $ km\,s$^{-1}$ and $ 205 < V_{\mathrm{T}} < 217 $ km\,s$^{-1}$, excluding stars with $ J-K\le 0.2 $ mag, using Gaussian smoothing with parameters $ \sigma_{V_T} = 0.4$, $ \sigma_{R}=4 $. The line of regression through the minima at constant distance is also shown. The minimum at the Solar radius is at $ V_{\mathrm{T}} = 210.8 \pm 0.4 $ km\,s$^{-1}$, based on Solar orbital velocity $ 225 $ km\,s$^{-1}$.}
	\label{Fig:20}
\end{figure}

To find the Solar motion relative to the LSR in the direction of Galactic rotation, I used Solar velocity $V_\odot = 225 \pm 5$ km\,s$^{-1}$, found from the measurement of the proper motion of Sgr A* (Reid and Brunthaller 2004), under the assumption that Sgr A* is stationary at the Galactic centre together with a combined estimate for $R_0$ of $ 7.4 \pm 0.2 $ kpc found from Reid (1993), Nishiyama et al. (2006), Bica et al. (2006), Eisenhauer et al. (2005), and Layden et al. (1996). The values of $R_0$ and $V_\odot$ are not critical to the calculation. The method for finding the LSR introduced by Francis and Anderson (2009a) fails for the RAVE population, because it yields a broad spread of minima from $ U_0 = -8 $km\,s$^{-1}$ to $ -16 $ km\,s$^{-1}$ and $ V_0 = -10 $ km\,s$^{-1}$ to $ -18 $ km\,s$^{-1}$. The circular speed curve is less clear than in the Hipparcos data, but it is visible as a minimum in the transverse velocity distribution for stars with close to circular motion for a greater range of Galactic radii. I plotted the transverse velocity distribution against distance to the Galactic centre for stars with $ | U + 8 | < 6 $ km\,s$^{-1}$, excluding stars with $ J-K\le 0.2$ mag, using Gaussian smoothing with parameters $ \sigma_{V_T} = 0.4$, $ \sigma_{V_R}=4 $ (figure 20). I found the line of regression through the minimum values. The range is chosen to exclude the peak of new star formation seen in figure 11. At the Solar radius this gives an estimate of the LSR, $ V_0 = 15.3 \pm 0.4 $ km\,s$^{-1}$, in close agreement with the value obtained from XHIP, $V_0 = 14.6 \pm 0.4 $ km\,s$^{-1}$. The line of regression has a slope $ -9.8 $ km\,s$^{-1}$ kpc$^{-1}$, significant at the 92\% level.

\subsection{RAVE and spiral structure}\label{sec:5.8}
For an elliptical orbit the eccentricity vector is defined as the vector pointing toward pericentre and with magnitude equal to the orbit's scalar eccentricity. It is given by
\begin{equation}\label{eq:5.8.1}
\mathbf{e}=\frac{|\mathbf{v}|^{2}\mathbf{r}}{\mu} - \frac{(\mathbf{r}\cdot\mathbf{v})\mathbf{v}}{\mu} - 
\frac{\mathbf{r}}{|\mathbf{r}|}.
\end{equation}
where $\mathbf{v}$ is the velocity vector, $\mathbf{r}$ is the radial vector, and $\mu=GM$ is the standard gravitational parameter for an orbit about a mass $M$, (e.g., Goldstein 1980; Arnold 1989). For a Keplerian orbit the eccentricity vector is a constant of the motion. Stellar orbits are not strictly elliptical, but rosette orbits can usefully be regarded as precessing ellipses and the eccentricity vector remains a useful measure (the Laplace-Runge-Lenz vector, which is the same up to a multiplicative factor, is also used to describe perturbations to elliptical orbits).
\begin{figure}
	\centering
		\includegraphics[width=0.47\textwidth]{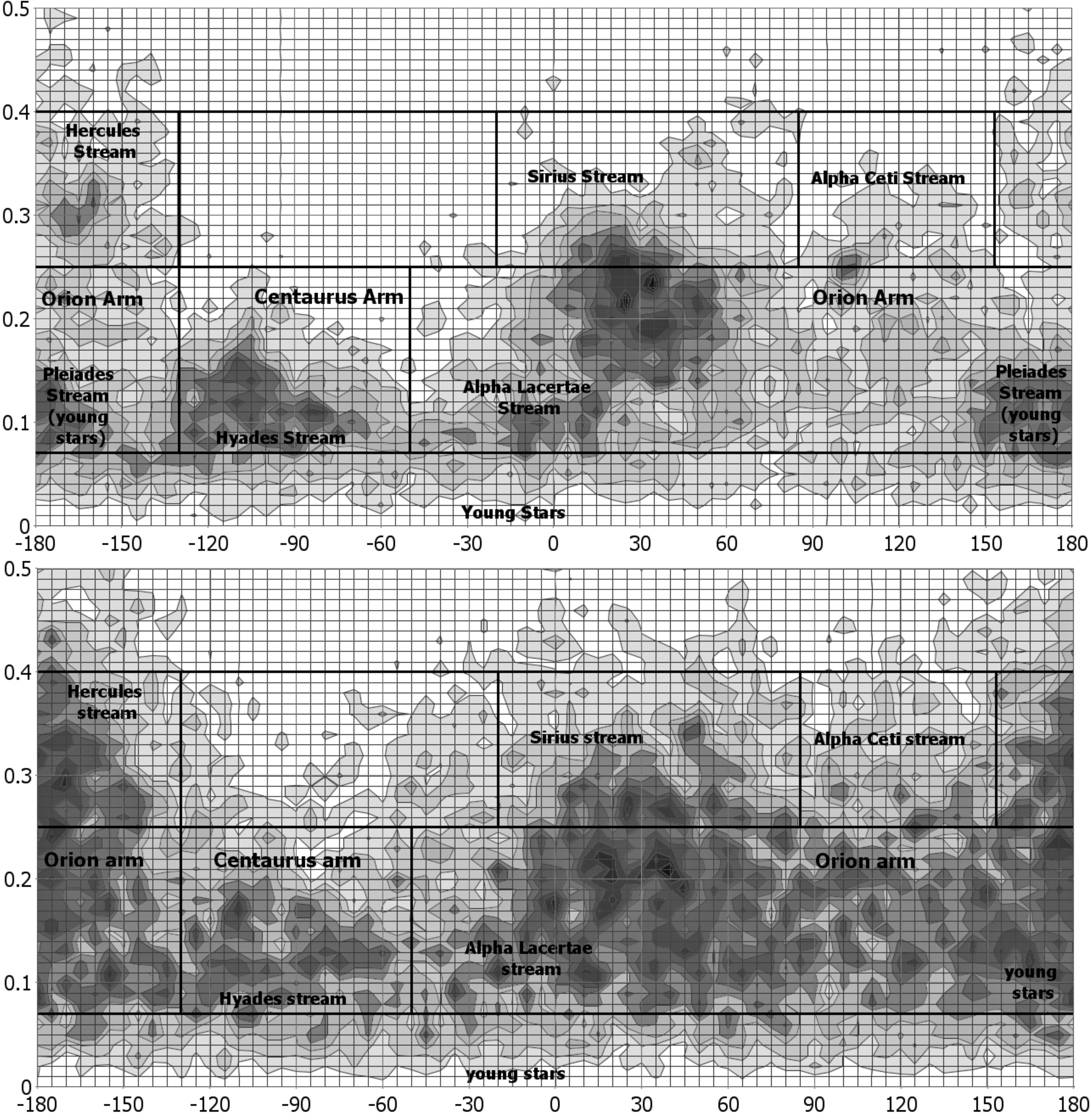}
		\caption{The $ e $-$ \phi $ distribution (eccentricity-true anomaly) using Gaussian smoothing for Hipparcos (top) and for RAVE (bottom). }
	\label{Fig:21}
\end{figure}
\begin{figure}
	\centering
		\includegraphics[width=0.47\textwidth]{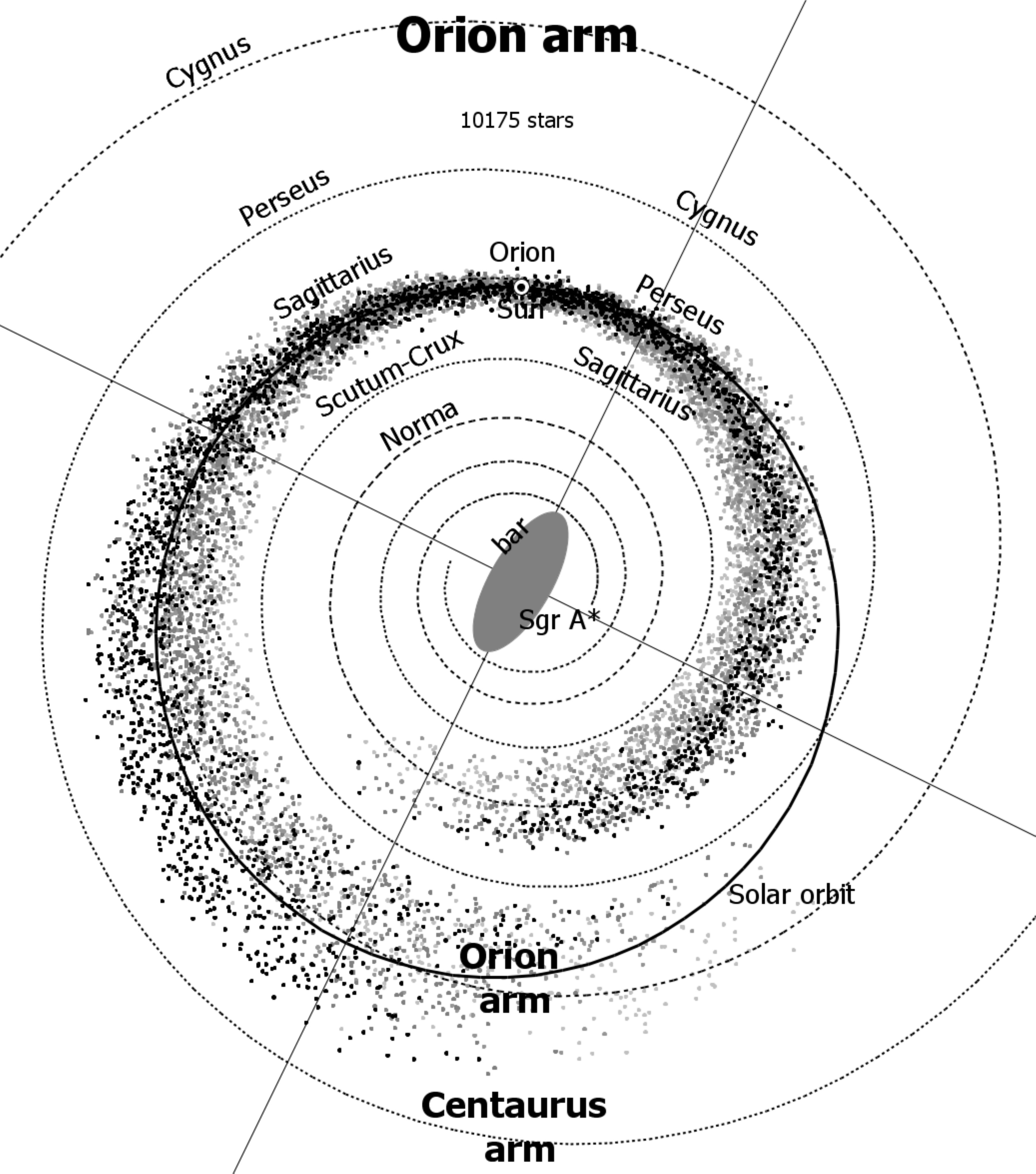}		\includegraphics[width=0.47\textwidth]{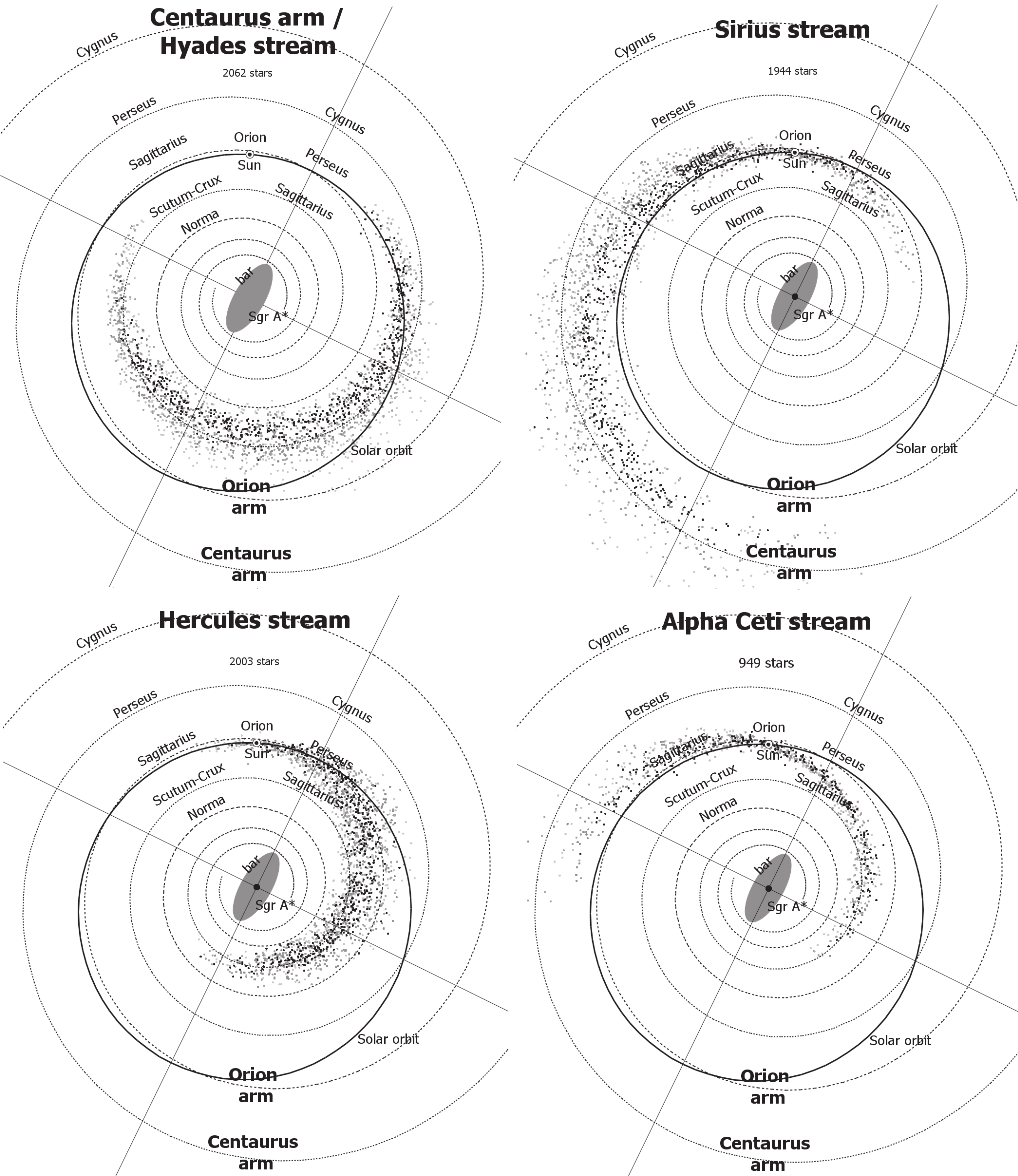}
		\caption{Two-armed logarithmic spiral with a pitch angle of $ 5.56 $\textdegree found from the distributions of neutral and ionised gas, giant molecular clouds, and the 2MASS stellar density (FA12), showing the Solar orbit (eccentricity 0.159), major axis and latus rectum. For each of five major kinematic groups RAVE stars are shown at a random position on the inward part of their orbits. Stars in with stream membership quality index 1, denoting the densest part of the stream, are shown in black, stars with quality indices 2 - 4 are have paler grey for a higher index. }
	\label{Fig:22}
\end{figure}
\begin{table}\label{(table6)}
\begin{tiny}
\begin{flushleft}
\begin{tabular}{lrrrrrr} 
\textbf{group}	 	 & $ e_{\mathrm{min}}	 $ & $ e_{\mathrm{max}}	 $ & $ \phi_{\mathrm{min}}	 $ & $ \phi_{\mathrm{max}}	 $ & Hip	 & RAVE\\
young stars	 & $ 0.0	 $ & $ 0.07	 $ & $ - $ & $ - $ & $ 12.8$\%	 & $ 9.3 $\% \\
Orion arm	 & $ 0.07	 $ & $ 0.25	 $ & $ -55$\textdegree & $ -130$\textdegree & $ 48.4\%	 $ & $ 46.4\%$\\
Centaurus-Hyades & $ 0.07	 $ & $ 0.25	 $ & $ -130$\textdegree & $ -55$\textdegree & $ 12.7\%	 $ & $ 9.4\%$\\
Hercules stream	 & $ 0.25	 $ & $ 0.4	 $ & $ 153$\textdegree & $ -130$\textdegree & $ 6.5\%	 $ & $ 9.1\%$\\
Sirius stream	 & $ 0.25	 $ & $ 0.4	 $ & $ -20$\textdegree & $ 85$\textdegree & $ 7.6$\%	 & $ 8.9$\%	 \\
Alpha Ceti stream & $ 0.25	 $ & $ 0.4	 $ & $ 85$\textdegree & $ 153$\textdegree & $ 3.1\%	 $ & $ 4.3$\%\\
high eccentricity & $ 0.4	 $ & $ -	 $ & $ - $ & $ -	 $ & $ 6.8\%	 $ & $ 9.0\%$\\
Total	 	 & $ 	 $ & $ 	 $ & $ 	 $ & $ 	 $ & $ 98.0\%	 $ & $ 96.5\%$\\
\end{tabular}
\caption{Statistics for populations in RAVE identified as dwarfs, T-Tauri and subgiants (TT\&S), protostars and red giant branch (PS\&R), and red clump giants (RC).}
\end{flushleft}
\end{tiny}
\end{table}

For a Galactic orbit, apart from local perturbations (including perturbations due to spiral arms) the gravitational force is directed toward the Galactic centre, and varies slowly with Galactic radius because of the mass distribution in the disc and halo. The equivalent central mass is the effective mass which would generate the same gravitational force on the star if all the mass generating the gravitational field was centrally placed. We can treat the orbit, over some time span, as a perturbation to the osculating Kepler orbit, which is the orbit it would have if the equivalent central mass were constant. The osculating Kepler orbit is an ellipse, characterised by current distance, $R$, to the Galactic centre, eccentricity, $ e=|\mathbf{e}| $, and the true anomaly, $\phi$, or angle between the eccentricity vector and the star subtended at the Galactic centre. Comparison of the eccentricity-true anomaly ($ e $-$\phi$) distributions for Hipparcos and RAVE (figure 21, with an adopted Solar orbital velocity of 225 km\,s$^{-1}$) emphasises the continuation of local streaming motions to distances around 500 pc, with some change to the Pleiades stream. The major kinematic groups are given in table 6. Bounding values of the true anomaly dividing the streams are chosen to match regions of low frequency in the RAVE distribution. The RAVE data suggest that $ \phi_\mathrm{min} = -130 $\textdegree is a better boundary between the Orion arm and the Hyades stream than $ \phi_\mathrm{min} = -140 $\textdegree, which is suggested by Hipparcos and was used in FA12. Other boundaries are unchanged. The boundaries in eccentricity are found from the alignments seen in figure 22, as described below.

The relative proportions of stars in each stream are very similar for RAVE and Hipparcos, but RAVE has more stars in the faster streams and in high eccentricity orbits, reflecting the fact the RAVE population is older, being further from the Galactic plane (restricting the height of RAVE stars above the Galactic plane reduces the proportions in these groups). RAVE has a higher proportion of Orion arm to Centaurus arm stars because the greater distances in RAVE incorporate a greater region of the Orion arm, including more stars from the denser central part of the arm. 

The alignment of orbits with spiral structure means that few local stars have true anomaly between $ -60 $\textdegree and $ -30 $\textdegree or between $ -140 $\textdegree and $ -120 $\textdegree, at which parts of their motion stars are travelling between the arms outward from the Galactic centre. Relatively few stars have true anomaly between $ 90 $\textdegree and $ 130 $\textdegree, at which part of their orbit stars are predicted to mainly lie on the outer part of a spiral arm, at greater than 500 pc from the Sun. The RAVE population has a larger proportions than Hipparcos with true anomaly between $ 50 $\textdegree and $ 150 $\textdegree, during which part of the orbit stars lie towards the central and outer parts of the arm, beyond the range of accurate Hipparcos measurement. 

I divided the stars in each segment defined by table 6 by quartiles of the density at the position of each star on the $ e $-$\phi $ distribution (figure 21), and assigned a quality index, 1 - 4, to each star. Thus, stars in a region with density greater than the upper quartile show the tightest adherence to stream motions, and are given quality index 1, while those in a region less dense than the lower quartile are least matched to stream motions and are given quality index 4. Since orbits are precessing ovals (i.e. rosettes) they can be extrapolated with reasonable accuracy for part of an orbit using an elliptical approximation. Transposing each star by a random amount, less than half an orbit, to a point on the inward part of the orbit shows the spiral structure of the Milky Way, as illustrated by RAVE stars (figure 22). The result exactly matches the finding for Hipparcos stars seen in FA12, showing that orbits are aligned with a logarithmic bisymmetric spiral with pitch angle $ 5.56 \pm 0.06 $ found from the distributions of neutral and ionised gas, and also seen in the observed distribution of peaks in the Galactic plane in data from 2MASS. Stars with the tightest adherence to stream motions (black) show the closest alignment with spiral structure.
\begin{figure}
	\centering
		\includegraphics[width=0.47\textwidth]{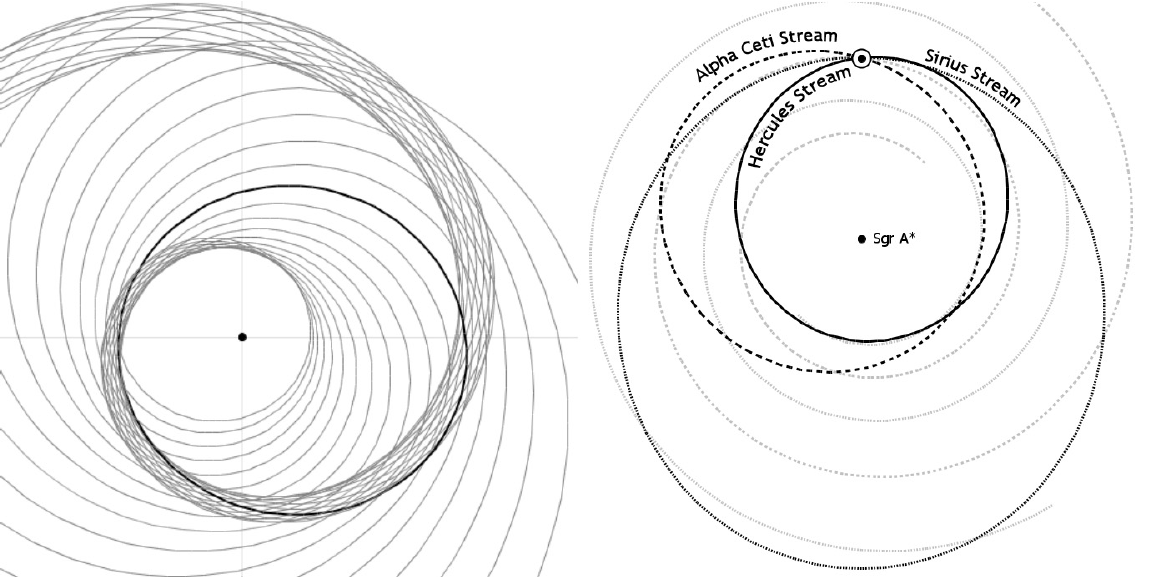}
		\caption{Left: equiangular spiral with a pitch angle of 11\textdegree, constructed by repeatedly enlarging an ellipse with eccentricity 0.3 by a factor 1.05 and rotating it through 15\textdegree with each enlargement. Ellipses with eccentricity greater than about 0.25 have more than half their circumference within the spiral region. Ellipses with eccentricity greater than about 0.35 produce probably too broad a spiral structure to model a spiral arm with this pitch angle, but give a good fit for spirals with greater pitch angle. Lower eccentricity ellipses produce a narrower spiral structure and/or a fit with spirals of lower pitch angle. Right: alignment of orbits with eccentricity 0.29 with a spiral of pitch angle 5.56\textdegree. These minor modes have apocentre and pericentre aligned with different arms. For orbits passing through the solar neighbourhood, the continuous line corresponds to the Hercules stream. The dashed line shows the Alpha Ceti stream, and the dotted line represents the Sirius stream.}
	\label{Fig:23}
\end{figure}

The alignment of stellar motions with the Orion arm and the Hyades stream in figure 22 can be understood by repeatedly enlarging an ellipse by a constant factor, $k$, centred at the focus and rotating it by a constant angle, $\tau$, with each enlargement (figure 23, left). In this description of material arms there is no winding problem because the spiral depends on the \textit{paths} of stellar orbits, not on orbital velocity. The pitch angle of the spiral depends on $k$ and $\tau$, not on the eccentricity of the ellipse, but, for a given pitch angle, ellipses with a range of eccentricities can be fitted to the spiral, depending on how narrow one wants to make the spiral and what proportion of the circumference of each ellipse one wants to lie within it. Higher eccentricity orbits fit spirals with higher pitch angles. Thus, stars move along an arm on the inward part of their orbits, leaving the arm soon after pericentre, crossing the other arm on the outward part and rejoining the original arm near to apocentre.

The Hyades stream consists of stars crossing the Orion arm on the outward part of their orbits. Transposing them to the inward part of their orbits shows the alignment of their orbits with the Centaurus arm. A small number of stars are seen between the solar position and the arm. These will join the arm shortly after apocentre. The apparent alignment of stars with quality index 1 just inside the expected line of the Centaurus arm can be accounted for by perturbations of the orbit due to the gravity of the arms.

Fast moving streams are explained as minor modes of alignment for orbits of greater eccentricity straddling the arms; the orbit aligns with one spiral arm close to apocentre and to the other in the region of pericentre. Figure 23 (right) shows three minor modes of alignment for stars in the solar neighbourhood, corresponding to the Sirius, Hercules and Alpha Ceti streams. Francis \& Anderson (2009a) found that a sharp change in velocity components at age $ 9 \pm 1 $ Gyr (previously seen by Quillen \& Garnett 2000) is caused by an increased membership of old stars in high eccentricity streams.

Clearly orbits are not perfectly elliptical, and the projected positions of stars will be subject to perturbation. However, for distinct populations of stars, the close alignment between elliptical orbits projected from their current position in the neighbourhood of the Sun, and current spiral structure, found in neutral and ionised gas, in star forming regions and in the 2MASS stellar density confirms that the overall effect of perturbations due to the gravity of the arms reinforces the alignment with the arms, as well as the stability and low pattern speed of the Milky Way spiral.

\section{Conclusions}\label{sec:6}
I have calculated a data base of distances for 52\,794 RAVE stars, incorporating 39\,514 dwarfs, 9\,057 red clump giants, 2\,262 protostars and red giant branch, and 1\,961 T-Tauri and subgiants, identified by position on the colour-surface gravity diagram and calibrated to a large sample of Hipparcos stars. I have given a detailed analysis of the types of bias which can affect such a calibration, and calculated appropriate corrections. I have used a magnitude limited calibration sample to minimise the Lutz-Kelker bias and the Malmquist bias, which is the source of greatest bias in the calibration of luminosity distance to parallax. In practice, I found that the standard correction for the Malmquist bias is not unreasonable (nor is it optimal) for red giants, but works poorly for hot dwarfs because their distribution in space is not uniform and in magnitude is not Gaussian.

Comparison with Hipparcos dwarfs with parallax errors less than 2\% shows that errors in luminosity distances due to the width of the main sequence are better than 20\%. For correctly identified giants errors are around 10\%. These can be taken as realistic error estimates for RAVE stars, because both populations consist of mainly thin disc stars with complete range of stellar type, composition and age. For stars with Galactic latitude $ |b| >9.7 $\textdegree, corrections to distances for reddening are typically less than 5\%, and errors in reddening contribute very little to overall error. These figures show notable improvement on the errors reported by Zwitter et al. (2010) and on errors found using the method of Bilir et al. (2008). For each class, the calibrated distances are systematically less than distances calculated by Zwitter et al. (2010) from Padova isochrones, but distances for dwarfs are greater than those found using the method of Bilir et al. (2008). Much of the difference for stars with substantially greater isochrone distances can be attributed to the misidentification of stellar class, which applies to a limited number of stars. Distances are systematically less than isochrone distances, even after the removal of stars showing differences in identification of stellar class.

The resulting velocity distribution has been compared to that found from Padova isochrones. Lower dispersion and a better correspondence with the Hipparcos velocity distribution are found in the thin disc using distances from this calibration. The velocity plot for dwarfs indicates the better accuracy of distances calibrated to Hipparcos, showing similar features but being more compact. The high value of $ |\overline{U}| $ km\,s$^{-1}$ found for giants suggests that distances calculated by both methods may be overstated (as there is no reason to think proper motions are systematically high). If so, Padova isochrones are more seriously affected than the Hipparcos calibration. This is also suggested by the higher velocity dispersions found using isochrone distances.

It was found that metallicity has very little effect on distances using $H$ magnitudes and $J-K$ colours. There is reason to think that identifying stellar class using only surface gravity and colour is more accurate than attempting a fit to isochrones from a range of parameters by the method of either Zwitter et al. (2010) or Burnett and Binney (2010). It is not possible to make direct comparison with distances of Burnett et al. (2011), using the method of Burnett and Binney (2010) as no data base is provided and the description is insufficient for replication. The claim to improve on other methods appears to be based on unconventional tests which are open to interpretation. The claim to have demonstrated the robustness of isochrones is refuted by the age of the Hyades, for which a wide range of isochrone ages up to 12 Gyr have been given. They report slightly greater distances than found by Zwitter et al. (2010) for dwarfs and slightly smaller distances for giants, but `slightly' is not quantified. As distances calibrated to Hipparcos are significantly less than those of Zwitter et al. (2010) for both dwarfs and giants, it does not appear that this method holds any advantage. The statement that half the RAVE population are giants conflicts with the determination of stellar class using $\log g$ and colour, from which we find that dwarfs and pre-main sequence stars outnumber giants by two to one. This causes one to suspect that a significant proportion of isochrone distances are erroneously large because of misidentifications in stellar class.

The bulk of distances for the dwarf population lie between 200 and 500 pc, and usefully extends the size of the Solar neighbourhood in which we can plot the velocity distribution. There is a close correspondence between the structure of the velocity distribution seen in RAVE stars and that previously seen from Hipparcos, as well as expected changes in the velocity distribution with distance from the Sun. Similar streaming motions are seen in the velocity distribution, excepting that the Pleiades stream does not appear at the same position and strength on the $U$-$V$ diagram, because, in each region, young stars belong to associations with similar motions deriving from the giant molecular clouds in which they form; the motions of these regions vary with position. 

A substantial proportion of the population of RAVE giants are thick disc stars (as determined by distance from the Galactic plane and motion perpendicular to the Galactic plane). Little structure is observed within the thick disc. In contrast to the distributions of thin disc giants and main sequence stars, which have a well at the position of circular motion, the populations containing pre-main sequence stars have their largest peak close to the LSR, illustrating the kinematics of new stars.

The estimate of the LSR from RAVE, $ (U_0, V_0, W_0) = (14.9 \pm 1.7, 15.3 \pm 0.4, 6.9 \pm 0.1) $ km\,s$^{-1}$ shows excellent agreement with the current best estimate from XHIP which uses an almost entirely distinct population of stars and gives $ (U_0, V_0, W_0) = (14.1 \pm 1.1, 14.6 \pm 0.4, 6.9 \pm 0.1) $ km\,s$^{-1}$.

Alignments of the velocity distribution with a bisymmetric spiral are as strong in the population of RAVE stars as they are in Hipparcos, but the RAVE distribution shows a greater spread of velocities for stars within the Orion arm, as is predicted by Francis and Anderson 2009b for stellar positions at greater depth within the arm, and slightly greater proportions of stars in fast moving streams, as expected due to the greater age of stars further from the Galactic plane.

\section*{Acknowledgements}
This publication makes use of data products from the Two Micron All Sky Survey, which is a joint project of the University of Massachusetts and the Infrared Processing and Analysis Center/California Institute of Technology, funded by the National Aeronautics and Space Administration and the National Science Foundation.

\appendix

\section{}
 Using the Taylor expansion, under reasonable conditions,
\begin{equation*}
\begin{split}
\overline{y}&= \int\limits_{-\infty}^{\infty}f(x)dx\\
&=\int\limits_{-\infty}^{\infty}(y(\overline{x}) +(x-\overline{x})y'(\overline{x}) + (x-\overline{x})^2\frac{y''(\overline{x})}{2!} + \ldots) f(x)dx\\
&= y(\overline{x}) + \frac{y''(\overline{x})}{2!}\int\limits_{-\infty}^{\infty} (x-\overline{x})^2 f(x)dx + \ldots\\
&= y(\overline{x}) + \frac{y''(\overline{x})}{2}\sigma^2 + \sum\limits_{n=3}^{\infty} \frac{y^{(n)}(\overline{x})}{n!}\mu_n
\end{split}\end{equation*}
\\
\textbf{Example 1:} (parallax distance). If $ y=1000x^{-1} $ then 
\begin{equation*}
\begin{split}
&y'=-1000x^{-2}.\\
&y''=2000x^{-3}.
\end{split}\end{equation*}
The first two terms are:
\begin{equation}
\overline{y} \approx \frac{1000}{\overline{x}}(1+\frac{\sigma^2}{\overline{x}^2})
\end{equation}
which is a reasonable approximation to equation (\ref{eq:2.3.3}).\\
\\
\textbf{Example 2:} (distance modulus). If 
\begin{equation*}
y=5\log\frac{1000}{x}-5= 10-5\log x
\end{equation*}
then 
\begin{equation*}
\begin{split}
&y'=-5x^{-1}.\\
&y''=5x^{-2}.
\end{split}\end{equation*}
The first two terms are:
\begin{equation}
\overline{y} \approx 5\log \frac{1000}{\overline{x}} -5 + 2.5\frac{\sigma^2}{\overline{x}^2}
\end{equation}
The second order term is a poor approximation to equation (\ref{eq:2.4.2}); the reason is that a log law amplifies higher moments of the probability distribution, so that the series in equation (\ref{eq:2.2.1}) converges rather slowly.

\onecolumn

\section{Data}
The first ten rows of the data base of distances for 52\,794 RAVE stars are shown in table B1 and table B2. The full data base is available online and at CDS.

\begin{table}\label{(table7)}

\begin{tabular}{lllrrrrr} 
\textbf{Name}&\textbf{RAVE}&\textbf{2MASS}&\textbf{RAdeg}&\textbf{DEdeg}&\textbf{GLON}&\textbf{GLAT}&\textbf{Rlum}\\
C0000004-534721 &J000000.5-534721&00000048-5347212&$  0.00204166 $ & $-53.78925000 $ & $319.03141 $ & $-61.70206 $ & $ 521 $ \\
C0000013-650017 &J000001.3-650017&00000133-6500170&$  0.00550000 $ & $-65.00475000 $ & $311.56982 $ & $-51.25252 $ & $ 767 $\\
C0000033-274614 &J000003.4-274615&00000338-2746145&$  0.01412500 $ & $-27.77072222 $ & $ 26.74911 $ & $-78.58791 $ & $ 364 $\\

C0000050-450729 &J000005.0-450730&00000501-4507297&$  0.02095833 $ & $-45.12494444 $ & $329.24721 $ & $-69.28896 $ & $ 396 $\\
T6415\textunderscore 00641\textunderscore 1  &J000009.7-251929&00000971-2519291&$  0.04045833 $ & $-25.32480556 $ & $ 38.96565 $ & $-78.36555 $ & $ 178 $\\
C0000114-462206 &J000011.4-462207&00001139-4622065&$  0.04750000 $ & $-46.36852778 $ & $327.32881 $ & $-68.25711 $ & $ 332 $\\
T7526\textunderscore01241\textunderscore1   &J000014.2-415524&00001415-4155242&$  0.05895833 $ & $-41.92338888 $ & $334.93342 $ & $-71.87645 $ & $ 147 $ \\
T7523\textunderscore00775\textunderscore1   &J000016.6-390110&00001656-3901099&$  0.06908333 $ & $-39.01941666 $ & $341.61223 $ & $-74.02527 $ & $ 100 $\\
T5260\textunderscore00889\textunderscore1   &J000018.6-094054&00001856-0940535&$  0.07737500 $ & $ -9.68152777 $ & $ 85.86146 $ & $-68.78911 $ & $ 410 $ \\
C0000202-520757 &J000020.3-520757&00002028-5207567&$  0.08445833 $ & $-52.13252777 $ & $320.46549 $ & $-63.22123 $ & $ 302 $ \\
\end{tabular}
\caption{Cols 1 to 8, first ten rows of dwarfs.dat. \textbf{Name}: Target designation; \textbf{RAVE}: RAVE designation (JHHMMSS.S+DDMMSS); \textbf{2MASS} 2MASS designation (Cat. II/246); \textbf{RAdeg}:  Right ascension (J2000.0, Ep=J2000); \textbf{DEdeg}: Declination (J2000.0, Ep=J2000) \textbf{GLON}:    	Galactic longitude; \textbf{GLAT}: Galactic latitude; \textbf{Rlum}:Luminosity distance. Data are taken from RAVE, except for Rlum}
\end{table}
\begin{table}\label{(table8)}
\begin{tabular}{rrrrrrrrrrrr} 
\textbf{pmRA}&\textbf{pmDE}&\textbf{RV}&\textbf{J}&\textbf{H}&\textbf{K}&\textbf{logg}&\textbf{[M/H]}&\textbf{[a/Fe]}&\textbf{band}&\textbf{Qual}&\textbf{MASK} \\
$  15.4 $ & $  -2.3 $ & $  21.0 $ & $11.831 $ & $11.547 $ & $11.489 $ & $3.87 $ & $  0.01 $ & $0.00$&g   &     &$1.00$\\
$  19.7 $ & $  -6.0 $ & $   8.9 $ & $12.057 $ & $11.864 $ & $11.800 $ & $     $ & $       $ & $  $&g   &     &$1.00$\\
$  13.0 $ & $ -12.9 $ & $  -9.0 $ & $11.683 $ & $11.327 $ & $11.263 $ & $     $ & $       $ &    &y   &     &$    $\\
$  -4.6 $ & $ -23.4 $ & $  55.6 $ & $11.483 $ & $11.180 $ & $11.103 $ & $     $ & $       $ &     &y   &     &$    $\\
$  -1.5 $ & $  -6.0 $ & $  -2.7 $ & $ 8.946 $ & $ 8.766 $ & $ 8.658 $ & $     $ & $       $ &     &g   &     &$    $\\
$  -0.6 $ & $ -19.3 $ & $  11.2 $ & $11.167 $ & $10.845 $ & $10.779 $ & $     $ & $       $ &     &y  &  & $ $\\
$  -1.3 $ & $   6.5 $ & $  11.9 $ & $ 8.658 $ & $ 8.472 $ & $ 8.351 $ & $4.27 $ & $ -0.38 $ & $0.35$&g   &     &$1.00$\\
$  26.6 $ & $ -10.9 $ & $   7.6 $ & $ 9.337 $ & $ 8.920 $ & $ 8.839 $ & $4.53 $ & $  0.07 $ & $0.00$&r   &     &$0.97$\\
$  -1.6 $ & $  -7.5 $ & $  -5.5 $ & $10.539 $ & $10.332 $ & $10.296 $ & $4.31 $ & $ -0.04 $ & $0.14$&g   &     &$1.00$\\
$ -12.0 $ & $   5.1 $ & $  30.6 $ & $10.851 $ & $10.549 $ & $10.473 $ & $4.17 $ & $ -0.26 $ & $0.20$&y   &     &$1.00$\\

\end{tabular}

\caption{Cols 9 to 20, first ten rows of dwarfs.dat. \textbf{pmRA}: ? proper motion RA; \textbf{pmDE}: ? proper motion DE; \textbf{RV}:      	Heliocentric radial velocity;	\textbf{J}: 2MASS with correction for reddening; \textbf{H}: 2MASS with correction for reddening;	\textbf{K}: 2MASS with correction for reddening		\textbf{logg}: ? Gravity; \textbf{[M/H]}; \textbf{[a/Fe]}; \textbf{band}: b, g, y, r, WD, RC, TT\&S, PS\&R; \textbf{Qual}:  Spectra quality flag;	\textbf{MASK}: MASK flag. Data are taken from RAVE, except for JHK}

\end{table}

\twocolumn

\label{lastpage}


\begin{thebibliography}{}
\bibitem[\protect\citeauthoryear{Anderson \& Francis}{2012}]{b1} Anderson E. \& Francis C., 2012 (`XHIP'), Astron. Lett., 5, 331, (CDS Catalog V/137, XHIP).
\bibitem[\protect\citeauthoryear{Antoja et al.}{2012}]{b2}Antoja T. et al., 2012, arXiv:1205.0546v1 
\bibitem[\protect\citeauthoryear{Arnold}{1989}]{b3}Arnold, V.I., 1989, Mathematical Methods of Classical Mechanics, 2nd ed., New York: Springer-Verlag. p. 38
\bibitem[\protect\citeauthoryear{Bahcall \& Soneira}{1980}]{b4} Bahcall J. N., Soneira R. M., 1980, ApJS, 44, 73
\bibitem[\protect\citeauthoryear{Bica et al.}{2006}]{b5}Bica E., Bonatto C., Barbuy B., Ortolani S., 2006, Globular cluster system and Milky Way properties revisited, A\&A, 450, 1, pp.105-115
\bibitem[\protect\citeauthoryear{Bilir et al.}{2008}]{b6}Bilir S., Karaali S., Ak S., Yaz E., Cabrera-Lavers A., Co\c{s}kuno\u{g}lu K. B., 2008, MNRAS, 390, 1569
\bibitem[\protect\citeauthoryear{Binney \& Merrifield}{1980}]{b7}Binney \& Merrifield, 1998, Galactic Astronomy, Princeton University Press
\bibitem[\protect\citeauthoryear{Bovy et al.}{2012}]{b8}Bovy J., Rix H-W., Hogg M.W., 2012, ApJ, 751, 131, doi:10.1088/0004-637X/751/2/131
\bibitem[\protect\citeauthoryear{Breddels et al.}{2010}]{b9}Breddels M.A., Smith M.C., Helmi A., et al., 2010, A\&A, 511, 16
\bibitem[\protect\citeauthoryear{Burnett et al.}{2011}]{b10}Burnett B., Binney J., Sharma S. et al., 2011, A\&A, 532, A113
\bibitem[\protect\citeauthoryear{Burnett \& Binney}{2010}]{b11}Burnett B. \& Binney J. 2010, MNRAS, 407, 339
\bibitem[\protect\citeauthoryear{Burstein \& Heiles}{1978}]{b12}Burstein D. \& Heiles C. 1978, ApJ, 225, 40
\bibitem[\protect\citeauthoryear{Burstein \& Heiles}{1982}]{b13}Burstein D. \& Heiles C. 1982, AJ, 87, 1165
\bibitem[\protect\citeauthoryear{Co\c{s}kuno\u{g}lu et al.}{1980}]{b14}Co\c{s}kuno\u{g}lu B., Ak S., Bilir S., et al., 2011, MNRAS, 412, 2, 1237-1245.
\bibitem[\protect\citeauthoryear{Dommanget \& Nys}{2000}]{b15}Dommanget J., Nys O., 2000. Catalogue of the Components of Double and Multiple Stars, Observations et Travaux 54, 5. CDS Catalog I/274
\bibitem[\protect\citeauthoryear{Cutri et al.}{2003}]{b16}Cutri R.M. et al., 2003, The 2MASS All-Sky Catalog of Point Sources, 2003yCat.2246....0C
\bibitem[\protect\citeauthoryear{Eisenhauer et al.}{2005}]{b17}Eisenhauer F., Genzel R., Alexander T., Abuter R., Paumard T., Ott T., Gilbert A., Gillessen S., Horrobin M., Trippe S., Bonnet H., Dumas C., Hubin N., Kaufer A., Kissler-Patig M., Monnet G., Str\"{o}bele S., Szeifert T., Eckart A., Sch\"{o}del R., Zucker S., 2005, SINFONI in the Galactic Center: Young Stars and Infrared Flares in the Central Light-Month, Astrophys. J., 628, 246-259
\bibitem[\protect\citeauthoryear{Francis \& Anderson}{2009a}]{b18}Francis C. \& Anderson E., 2009a, New Astron., 14, pp. 615-629
\bibitem[\protect\citeauthoryear{Francis \& Anderson}{2009b}]{b19}Francis C. \& Anderson E., 2009b, Proc. R. Soc. A, 465, 3401-3423
\bibitem[\protect\citeauthoryear{Francis \& Anderson}{2012}]{b20}Francis C. \& Anderson E., 2012, `FA12', MNRAS, 422, 2, 1283-1293
\bibitem[\protect\citeauthoryear{Goldstein}{1980}]{b21}Goldstein H., 1980, Classical Mechanics (2nd ed.). Addison Wesley. pp. 102-105, 421-422. 
\bibitem[\protect\citeauthoryear{Layden et al.}{1996}]{b22}Layden A. C., Hanson R. B., Hawley S. L., Klemola A. R., \& Hanley C. J., 1996, The Absolute Magnitude and Kinematics of RR Lyrae Stars Via Statistical Parallax. A.J., 112, 2110-2131
\bibitem[\protect\citeauthoryear{Haywood Smith}{2003}]{b23}Haywood Smith Jr., 2003, MNRAS, 338, 891-902
\bibitem[\protect\citeauthoryear{Holmberg, Nordstr\"om \& Andersen }{2007}]{b24}Holmberg J., Nordstr\"om B. Andersen J., 2007, A\&A, 475, 519-537
\bibitem[\protect\citeauthoryear{van Leeuwen}{2007}]{b25}van Leeuwen F., 2007, Springer, (CDS Catalog: I/311)
\bibitem[\protect\citeauthoryear{Kippenhahn \& Weigart}{1990}]{b25a}Kippenhahn R. \& Weigart A., 1990, Stellar Structure and Evolution, Springer-Verlag, Berlin
\bibitem[\protect\citeauthoryear{Lindeberg}{1922}]{b26}Lindeberg J. W., 1922, Mathematische Zeitschrift 15 (1): 211-225. doi:10.1007/BF01494395
\bibitem[\protect\citeauthoryear{Lutz \& Kelker}{1973}]{b27}Lutz T.E., Kelker D.H., 1973, PASP, 85, 573
\bibitem[\protect\citeauthoryear{Lutz \& Kelker}{1974}]{b28}Lutz T.E., Kelker D.H., 1974, BAAS, 6, 227
\bibitem[\protect\citeauthoryear{Lutz \& Kelker}{1975}]{b29}Lutz T.E., Kelker D.H., 1975, PASP, 87, 617
\bibitem[\protect\citeauthoryear{Malmquist}{1922}]{b30}Malmquist G., 1920, Jour. Medd. Lund Astron. Obs. Ser. II, 22:1?39
\bibitem[\protect\citeauthoryear{Marshall et al.}{1980}]{b32}Marshall D. J., Robin A. C., Reyl\'{e} C., Schultheis M., Picaud S., 2006, A\&A, 453, 635
\bibitem[\protect\citeauthoryear{Mason et al.}{2001-2010}]{b33}Mason B.D., Wycoff G.L., Hartkopf W.I., Douglass G.G., Worley C.E., 2001-2010. The Washington Visual Double Star Catalog, AJ, 122, 3466. CDS Catalog: B/wds
\bibitem[\protect\citeauthoryear{Nishiyama et al.}{2006}]{b34}Nishiyama S., Nagata T., Sato S., Kato D., Nagayama T., Kusakabe N., Matsunaga N., Naoi T., Sugitani K., and Tamura M., 2006, ApJ, 647, pt 1 pp. 1093-1098
\bibitem[\protect\citeauthoryear{Perryman et al.}{1997}]{b35}Perryman M.A.C. et al., 1997, ESA SP-1200, (CDS Catalog I/239)
\bibitem[\protect\citeauthoryear{Perryman et al.}{1998}]{b36}Perryman M.A.C., Brown A.G.A., Lebreton Y., et al., 1998, A\&A, 331, 81
\bibitem[\protect\citeauthoryear{Prialnik}{2010}]{b36a}Prialnik D., 2010, An Introduction to the Theory of Stellar Structure and Evolution, 2nd ed. Cambridge University Press
\bibitem[\protect\citeauthoryear{Quillen \& Garnett}{2000}]{b21a}Quillen A.C. \& Garnett D.R., 2000, astro-ph/0004210v3
\bibitem[\protect\citeauthoryear{Reid}{1993}]{b37}Reid M.J., 1993, ARA\&A, 31, 345-372
\bibitem[\protect\citeauthoryear{Reid \& Brunthaller}{2004}]{b38}Reid M.J. \& Brunthaller A., 2004, ApJ, 616, 872-884
\bibitem[\protect\citeauthoryear{Robin et al.}{2003}]{b39}Robin A. C., Reyl\'{e} C., Derri\`{e}re S., Picaud S., 2003, A\&A, 409, 523
\bibitem[\protect\citeauthoryear{Schlegel, Finkbeiner \& Davis}{1998}]{b40}Schlegel D. J., Finkbeiner D. P., Davis M., 1998, ApJ, 500, 525
\bibitem[\protect\citeauthoryear{Siebert et al.}{201}]{b41}Siebert A., Williams M.E.K., Siviero A., et al., 2011, A.J., 141, 187 
\bibitem[\protect\citeauthoryear{Skrutskie et al.}{2006}]{b44}Skrutskie M.F., Cutri R.M., Stiening R., et al., 2006, AJ, 131, 1163
\bibitem[\protect\citeauthoryear{Trumpler \& Weaver}{1953}]{b45}Trumpler R.J. \& Weaver H.F., 1953, Statistical Astronomy, Berkeley University Press, p.369
\bibitem[\protect\citeauthoryear{Zwitter, Siebert \& Munari}{2008}]{b46} Zwitter T., Siebert A., Munari U., 2008, A.J. 136 421-451
\bibitem[\protect\citeauthoryear{ Zwitter, Matijevi\v{c} \& Breddels}{2010}]{b47} Zwitter T., Matijevi\v{c} G., Breddels M.A., 2010, A\&A, 522, A54
\end{thebibliography}
\end{document}